\documentclass{article}

\usepackage{PRIMEarxiv}

\usepackage[utf8]{inputenc} % allow utf-8 input
\usepackage[T1]{fontenc}    % use 8-bit T1 fonts
\usepackage{hyperref}       % hyperlinks
\usepackage{url}            % simple URL typesetting
\usepackage{booktabs}       % professional-quality tables
\usepackage{amsfonts}       % blackboard math symbols
\usepackage{nicefrac}       % compact symbols for 1/2, etc.
\usepackage{microtype}      % microtypography
\usepackage{lipsum}
\usepackage{amsmath} % make sure this is in your preamble
\usepackage{cleveref}       % for \Cref cross-referencing
\usepackage{multirow}
\usepackage{epstopdf}
\usepackage{algorithmic}
\usepackage{fancyhdr}       % header
\usepackage{graphicx}       % graphics

\graphicspath{{media/}}     % organize your images and other figures under media/ folder

\ifpdf
  \DeclareGraphicsExtensions{.eps,.pdf,.png,.jpg}
\else
  \DeclareGraphicsExtensions{.eps}
\fi

\usepackage{enumitem}
\setlist[enumerate]{leftmargin=.5in}
\setlist[itemize]{leftmargin=.5in}

\title{From Features to States: Data–Driven Selection of Measured State Variables via RFE–DMDc}

\author{
  Haoyu Wang\thanks{Corresponding Author}, Roberto Ponciroli, Richard B. Vilim \\
  Nuclear Science and Engineering Division \\
  Argonne National Laboratory \\
  Lemont, IL, USA\\
  \texttt{\{haoyuwang, rponciroli, rvilim\}anl.gov} \\
   \And
  Andrea Alfonsi \\
  NuCube Energy, INC \\
  Pasadena, CA, USA\\
  \texttt{aalfonsi@nucube.energy} \\
}

\begin{document}
\maketitle

\begin{abstract}
The behavior of a dynamical system under a given set of inputs can be captured by tracking the response of an optimal subset of process variables (\textit{state variables}). For many engineering systems, however, first-principles, model-based identification is impractical, motivating data-driven approaches for Digital Twins used in control and diagnostics. In this paper, we present RFE-DMDc, a supervised, data-driven workflow that uses Recursive Feature Elimination (RFE) to select a minimal, physically meaningful set of variables to monitor and then derives a linear state-space model via Dynamic Mode Decomposition with Control (DMDc). The workflow includes a cross-subsystem selection step that mitigates feature \textit{overshadowing} in multi-component systems. To corroborate the results, we implement a GA-DMDc baseline that jointly optimizes the state set and model fit under a common accuracy cost on states and outputs. Across a truth-known RLC benchmark and a realistic Integrated Energy System (IES) with multiple thermally coupled components and thousands of candidate variables, RFE-DMDc consistently recovers compact state sets (\(\approx 10\) variables) that achieve test errors comparable to GA-DMDc while requiring an order of magnitude less computational time. The selected variables retain clear physical interpretation across subsystems, and the resulting models demonstrate competitive predictive accuracy, computational efficiency, and robustness to overfitting.
\end{abstract}

% keywords can be removed
\keywords{State Variables \and Recursive Feature Elimination \and Genetic Algorithm \and Dynamic Mode Decomposition with Control}

\section{Introduction}
\label{sec:Introduction}
In control theory and physics, \emph{state variables} are the minimal set of internal coordinates that fully describe a system’s dynamics and enable prediction of its response under specified inputs. Equivalently, the state vector is the lowest-dimensional Markov-sufficient representation: given the current state and inputs, it renders the dynamics Markovian and permits prediction of all subsequent behavior (\emph{minimal Markov set}). State coordinates are not unique—any invertible linear change of basis (similarity transform) produces an equivalent state—and they may be directly measured, partially measured, or estimated \cite{rugh1996linear}.\\
To set terminology, it is useful to distinguish \emph{state variables} from \emph{latent variables}. Latent variables are unobserved coordinates introduced for modeling convenience (e.g., to capture variance, improve prediction, or regularize), with no requirement of physical interpretability, minimality, or uniqueness under reparameterization. They often represent hidden factors influencing processes or measurements. Crucially, latent variables do not inherently ensure the Markov property for the dynamics unless this is explicitly enforced by the model \cite{bartholomew2011latent}. In short, state variables provide Markov sufficiency (typically minimal) for dynamical prediction, whereas latent variables are flexible internal representations for data modeling with no built-in Markov guarantee.\\
In the standard form, the dynamics of a physical system can be described by a set of $n$ coupled first-order Ordinary Differential Equations (ODEs), known as the state equations, in which the time derivative of each state variable is expressed in terms of state variables $\mathbf{x}(t)=[x_1(t), … ,x_n(t)]^T$ and input variables $\mathbf{v}(t)=[v_1(t), … ,v_m(t)]^T$ (\Cref{eq:std_form_x}).
\begin{equation}
\label{eq:std_form_x}
\begin{cases}
\begin{aligned}
\dot{x}_1 &= f_1(\mathbf{x}(t), \mathbf{v}(t), t), \\
\dot{x}_2 &= f_2(\mathbf{x}(t), \mathbf{v}(t), t), \\
\vdots \\
\dot{x}_n &= f_n(\mathbf{x}(t), \mathbf{v}(t), t)
\end{aligned}
\end{cases}
\end{equation}
For linear time-invariant (LTI) systems, \Cref{eq:std_form_x} reduces to the familiar state-space representation comprising a set of ODEs with constant coefficients (\Cref{eq:LTI_compact}). Output variables $\mathbf{y}(t)=[y_1(t), … ,y_p(t)]^T$ are expressed by linear combinations of state and input variables. This representation underpins most modern control and estimation algorithms. 
    \begin{equation}
    \label{eq:LTI_compact}
    \dot{\mathbf{x}}(t) = A\,\mathbf{x}(t) + B\,\mathbf{v}(t), 
    \qquad 
    \mathbf{y}(t) = C\,\mathbf{x}(t) + D\,\mathbf{v}(t),
    \end{equation}    
with constant matrices $A\!\in\!\mathbb{R}^{n\times n}$, $B\!\in\!\mathbb{R}^{n\times m}$, $C\!\in\!\mathbb{R}^{p\times n}$, and $D\!\in\!\mathbb{R}^{p\times m}$.\\
Data-driven identification of state-space models is a mature field. For LTI systems, subspace and realization methods (e.g., ERA/OKID, MOESP, N4SID, and instrumental-variable variants) recover latent state coordinates from input–output data with provable guarantees under excitation and noise assumptions \cite{van2012subspace,Ljung1998-em}. While these approaches yield accurate predictors, the resulting states are typically not tied to physically measured variables, which limits interpretability and complicates sensor planning. Nonlinear identification adds expressive power via polynomial and neural parametrizations (e.g., PNLSS, sparse identification of nonlinear dynamics, and neural ODEs) \cite{brunton2016sindy,pintelon2012system,chen2018neuralode}. In parallel, Koopman-based methods approximate nonlinear dynamics with a lifted linear model in an expanded observable space. Extended/DMD with control (EDMD/DMDc) and their kernelized variants learn linear dynamics for chosen dictionaries while handling actuation \cite{proctorDMDc,williams2015edmd,korda2018kcdc}. These models are powerful surrogates, but their internal coordinates (\emph{lifted observables}) are again not generally equal to measured process variables. A complementary line of work focuses on variable and sensor selection for high-dimensional dynamical systems. Column-pivoted QR and leverage-score sampling provide greedy yet effective subsets of measurements for regression and state estimation \cite{manohar2018data}. Optimization-based formulations using sparsity, mutual information, or Fisher information link selection to prediction or estimation performance \cite{joshi2009sensor,krause2008near}. In practice, feature-ranking schemes such as Recursive Feature Elimination (RFE) are popular because they couple naturally with user-defined performance metrics and return a fixed-size set of top-ranked variables from a large candidate pool \cite{GUYONrfe,sauerbrei2020state}.\\
In this work, we target a distinct goal: constructing a minimal Markov state from \emph{physically meaningful process variables}, selecting a small interpretable subset that directly serves as the LTI surrogate state for control and diagnostic tasks. We operationalize this with an RFE wrapper around Dynamic Mode Decomposition with Control \cite{proctorDMDc} (RFE--DMDc), using a supervised accuracy cost defined over states and outputs across multiple operating conditions. The approach actively interrogates the system across multiple operating conditions; the selected state set is required to be \emph{common} to all realizations. To guard against biases that can arise in multi-component systems (e.g., when variables from a high-gain subsystem overshadow others during ranking), we introduce a \emph{cross-subsystem} selection step that balances variable importance across components before the final elimination phase. To corroborate the selections and provide a compute–accuracy reference, we implement a GA--DMDc baseline that jointly optimizes the state set and the DMDc fit under a common accuracy cost defined on states and outputs. As representative test cases, we consider (i) a truth-known RLC circuit to illustrate identifiability under controlled conditions, and (ii) an Integrated Energy System (IES) comprising multiple thermally coupled generation and storage units, a setting in which first-principles state selection is rarely obvious.\\
The paper is organized as follows. In \Cref{sec:DMDc}, a brief overview of the DMDc algorithm is provided. \Cref{sec:GA_DMDc} and \Cref{sec:RFE_DMDc} describe the GA-informed and RFE-based selection procedures, respectively, including the cross-subsystem step. \Cref{sec:assessment} details the test cases. \Cref{sec:results} reports selection outcomes and the computational efficiency. Finally, \Cref{sec:conclusion} summarizes findings and outlines future work.      
\section{Brief Overview of the Dynamic Mode Decomposition with Control algorithm}
\label{sec:DMDc}
DMDc is a data-driven identification method that fits a \emph{discrete-time} state-space model from time series of measured states and inputs. 
Given sampled trajectories \(\mathbf{v}(k)\) and \(\mathbf{x}(k)\) at step \(\Delta t\), DMDc estimates the matrices \((A^{d},B^{d})\) in \Cref{eq:discrete_LTI_x}.
\begin{equation}
\label{eq:discrete_LTI_x}
\mathbf{x}(k+1) \;=\; A^{d}\,\mathbf{x}(k) + B^{d}\,\mathbf{v}(k).
\end{equation}
When a continuous-time LTI surrogate \((A,B,C,D)\) is also considered (\Cref{eq:LTI_compact}), the discrete and continuous realizations are related under a zero-order hold (inputs held constant over each sampling interval) (\Cref{eq:discretized_matrices}).
\begin{equation}
\label{eq:discretized_matrices}
A^{d} = e^{A\Delta t}, 
\qquad 
B^{d} = \Big(\int_{0}^{\Delta t} e^{A\tau}\, \mathrm{d}\tau\Big) B,
\qquad 
C^{d}=C,\quad D^{d}=D.
\end{equation}
To construct the regression, we stack time-shifted snapshots of states and inputs (each column is one time step) for \(k\in\{1,\ldots,l-1\}\), as shown in \Cref{eq:dmdc_eq_1}. We use uniform sampling throughout; if \(\Delta t\) is nonuniform, a time-varying formulation is required rather than standard DMDc.
\begin{equation}
\label{eq:dmdc_eq_1}
\begin{aligned}
& \mathrm{X}^{\prime}=\left[\begin{array}{cccc}\mid & \mid & & \mid \\ \mathbf{x}(2) & \mathbf{x}(3) & \cdots & \mathbf{x}(l) \\ \mid & \mid & & \mid\end{array}\right] \\ 
& \mathrm{X}=\left[\begin{array}{cccc}\mid & \mid & & \mid \\ \mathbf{x}(1) & \mathbf{x}(2) & \cdots & \mathbf{x}(l-1) \\ \mid & \mid & & \mid\end{array}\right] \\ 
& \mathrm{V}=\left[\begin{array}{cccc}\mid & \mid & & \mid \\ \mathbf{v}(1) & \mathbf{v}(2) & \cdots & \mathbf{v}(l-1) \\ \mid & \mid & & \mid\end{array}\right] 
\end{aligned}
\end{equation}
The relation in \Cref{eq:discrete_LTI_x} can be expressed in matrix form to incorporate the defined data matrices. As the state-space matrices are obtained through a best-fit process using the training dataset, the relation is only approximate (\Cref{eq:dmdc_eq_2}).  
\begin{align}
\label{eq:dmdc_eq_2}
\mathrm{X}^{\prime} &\approx \hat{A}^d\mathrm{X} + \hat{B}^d\mathrm{V}
\end{align}
Since there are two unknown matrices, it can be rewritten as shown in \Cref{eq:dmdc_eq_3}:
    \begin{align}
    \label{eq:dmdc_eq_3}
    \mathrm{X}^{\prime} \approx {G}\mathrm{\Omega}
    \end{align}
where ${G}=[\hat{A}^d,\hat{B}^d$], and $\Omega=[\mathrm{X},\mathrm{V}]$. The operator ${G}\in \mathbb{R}^{n\times(n+m)}$ can be calculated as:
    \begin{align}
    \label{eq:dmdc_eq_4}
    \mathit{G}=\mathrm{X}^{\prime} \mathrm{\Omega}^\dagger
    \end{align}
where the $^\dagger$ indicates the pseudoinverse. To solve for the $\Omega^\dagger$ in \Cref{eq:dmdc_eq_4} in a computationally efficient way, the Singular Value Decomposition (SVD) is applied to the augmented data matrix $\mathrm{\Omega}$ (\Cref{eq:dmdc_eq_5}).
    \begin{align}
    \label{eq:dmdc_eq_5}    \mathrm{\Omega}=\mathrm{U}\mathrm{\Sigma}\mathrm{W}^{*}\approx\tilde{\mathrm{U}}\tilde{\mathrm{\Sigma}}\tilde{\mathrm{W}}^{*}
    \end{align}
where $\tilde{\mathrm{U}}\in\mathbb{R}^{(n+m)\times{q}}$, $\tilde{\mathrm{\Sigma}}\in\mathbb{R}^{q\times{q}}$, and $\tilde{\mathrm{W}}\in\mathbb{R}^{(l-1)\times{q}}$ are the truncated SVD components, and $^{*}$ denotes the complex conjugate transpose. The truncation value $q$ is the number of non-zero elements in $\Sigma$. In this work, the truncation value $q$ is set to limit the condition number of $\mathrm{\Sigma}$ matrix to below $10^9$. \Cref{eq:dmdc_eq_6} provides an approximation of $G$:
    \begin{align}
    \label{eq:dmdc_eq_6}
    \mathit{G}\approx\mathrm{X}^{\prime} \tilde{\mathrm{W}}\tilde{\mathrm{\Sigma}}^{-1}\tilde{\mathrm{U}}^{*}
    \end{align}

By breaking the linear operator $\tilde{\mathrm{U}}$ into two separate components according to the dimensions of $\mathbf{x}(k)$ and $\mathbf{v}(k)$, the approximations of the matrices $\hat{A}^d$ and $\hat{B}^d$ can be found (\Cref{eq:dmdc_eq_7}).
    \begin{align}
    \label{eq:dmdc_eq_7}
    [\hat{A}^d,\hat{B}^d]\approx[\mathrm{X}^{\prime} \tilde{\mathrm{W}}\tilde{\mathrm{\Sigma}}^{-1}\tilde{\mathrm{U}}_1^{*},\mathrm{X}^{\prime} \tilde{\mathrm{W}}\tilde{\mathrm{\Sigma}}^{-1}\tilde{\mathrm{U}}_2^{*}]
    \end{align}
where $\tilde{\mathrm{U}}_1\in\mathbb{R}^{n\times{q}}$ and $\tilde{\mathrm{U}}_2\in\mathbb{R}^{m\times{q}}$.

To complete the discrete-time state-space model, we estimate the output maps \(C^{d}\) and \(D^{d}\). We set \(D^{d}=0\) because there is no direct (zero-lag) feedthrough from inputs to outputs over one sampling interval. Given training snapshots of states and outputs (see \Cref{eq:dmdc_eq_9} for \(Y\)), we can fit \(C^{d}\) by least squares.
\begin{equation}
\label{eq:dmdc_eq_9}
\mathrm{Y}=\left[\begin{array}{cccc}\mid & \mid & & \mid \\ \mathbf{y}(1) & \mathbf{y}(2) & \cdots & \mathbf{y}(l-1) \\ \mid & \mid & & \mid\end{array}\right] 
\end{equation}
The output equation in the state-space model takes the matrix form $Y={C}^{d}X$. The minimum-norm Moore–Penrose solution \cite{moore_inverse} is reported in \Cref{eq:dmdc_C_fit}.
\begin{equation}
\label{eq:dmdc_C_fit}
\hat{C}^{d} \;=\; \arg\min_{C}\,\|Y - C X\|_{F}^{2} \;=\; Y\,X^{\dagger},
\qquad\text{so that}\qquad
Y \approx \hat{C}^{d}X.
\end{equation}
    
\section{Cost-Function–Driven GA--DMDc for State Selection}
\label{sec:GA_DMDc}
\begin{figure}[htbp]
  \centering
  \includegraphics[width=0.4\textwidth,clip,keepaspectratio]{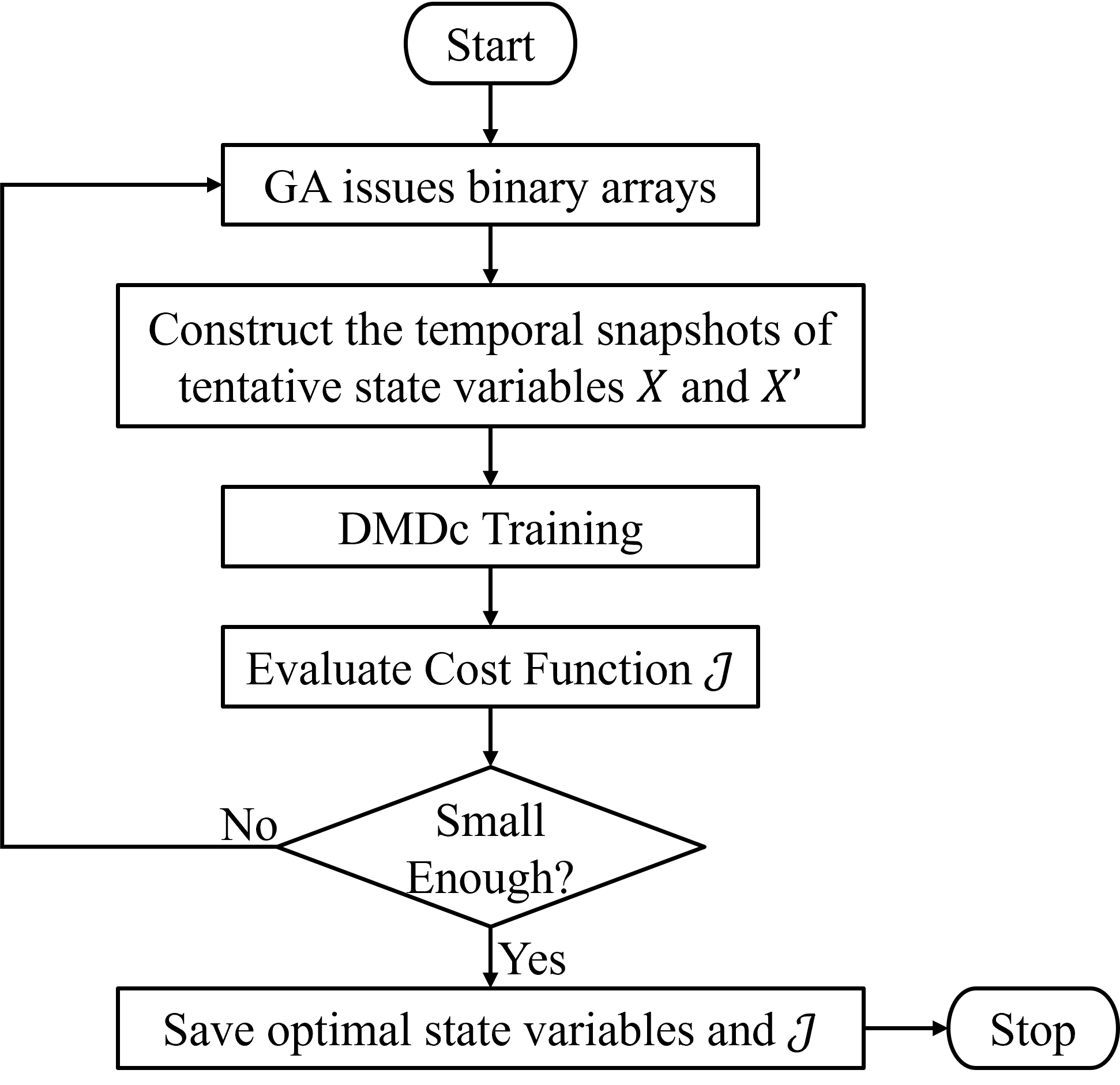}
  \caption{Flowchart of the GA--DMDc baseline used for comparison with RFE--DMDc.}
  \label{fig:GA}
\end{figure}

We adopt a three-step procedure to select a representative set of measured variables to serve as state variables. First, the multi-perturbation dataset is partitioned into training and testing subsets. On the training data, a genetic algorithm (GA) selects candidate state sets by minimizing a supervised accuracy cost, with each candidate evaluated via a DMDc fit. We then fit discrete-time state-space matrices with DMDc on the training set using the selected states. Finally, the predicted trajectories of the selected state variables $\mathbf{\hat{x}}(k\geq1)$ and output variables $\mathbf{\hat{y}}(k\geq1)$ are calculated by using the trained matrices, the input trajectory $\mathbf{v}(k\geq0)$, and the initial conditions $\mathbf{\hat{x}}(k=0)$(\Cref{eq:discrete_LTI_1}, \Cref{eq:discrete_LTI_2}, \Cref{eq:discrete_LTI_3}), and compared against ground truth, i.e., the expected trajectories in the testing data-set.
\begin{align}
\label{eq:discrete_LTI_1}
\mathbf{\hat{x}}(0) &= \mathbf{x}(0) 
\end{align} 
\begin{align}
\label{eq:discrete_LTI_2}
\mathbf{\hat{x}}(k) &= \hat{A}^d\mathbf{\hat{x}}(k-1)+\hat{B}^d\mathbf{v}(k-1), k\geq1
\end{align}    
\begin{align}
\label{eq:discrete_LTI_3}
\mathbf{\hat{y}}(k) &= \hat{C}^d\mathbf{\hat{x}}(k), k\geq1
\end{align}
The cost function accounting for the accuracy of both state variable and output variable predictions is defined as follows. Let \(n\) be the number of selected states and \(p\) the number of outputs.
Channel-wise scales \(\sigma(x_i)\) and \(\sigma(y_j)\) are computed \emph{on the training set only} and reused unchanged on the test set. For a rollout of length \(l\), we evaluate \Cref{eq:cost_function}.
\begin{equation}
\label{eq:cost_function}
\mathcal{J} \;=\; 
\frac{1}{n\,l}\sum_{i=1}^{n}\sum_{k=1}^{l}\!\left(\frac{\hat{x}_i(k)-x_i(k)}{\sigma(x_i)}\right)^{\!2}
\;+\;
\frac{1}{p\,l}\sum_{j=1}^{p}\sum_{k=1}^{l}\!\left(\frac{\hat{y}_j(k)-y_j(k)}{\sigma(y_j)}\right)^{\!2}
\end{equation}
This normalized mean-squared error (MSE) estimates the average discrepancy on both states and outputs. Importantly, \(\mathcal{J}\) does not necessarily decrease as more variables are selected, which provides an implicit pressure toward compact state sets, as shown in the later \Cref{sec:results}. The same metric is used consistently to compare RFE--DMDc and GA--DMDc selections on the test set.\\
A DMDc-informed GA scheme (GA--DMDc) was also developed to provide a baseline for the results of the RFE-based algorithm. Each GA chromosome is a binary mask over the candidate variables. Given a mask, we form the corresponding state snapshot matrices \((\mathrm{X},\mathrm{X}',\mathrm{V})\), fit \((\hat{A}^d,\hat{B}^d,\hat{C}^d)\) via DMDc on the training set, and score the candidate by \(\mathcal{J}\). The GA minimizes \(\mathcal{J}\) over masks subject to a user-specified upper bound on the state count (the number of selected variables is chosen implicitly). In \Cref{fig:GA}, the steps of GA--DMDc scheme are represented. At the beginning of each iteration, an array of binary numbers, whose length is equal to the number of the candidate state variables, is issued. Temporal snapshots of tentative state variables ($\mathrm{X}^{\prime}$ and $\mathrm{X}$ in \Cref{eq:dmdc_eq_1}) participate in the DMDc training and lead to a tentative set of state-space representation matrices. Small values of the cost function indicate that the candidate state variables provide a sufficiently accurate characterization of the system dynamics, and then the GA will stop; otherwise, the GA will alter the binary arrays by starting another iteration of the optimization process.

Key GA settings are summarized in \Cref{tab:GA_parameters}. The \emph{Population Size} controls exploration (larger values generally improve accuracy at increased cost); we explored values from 120 to 3,840 and used 480 by default. Other parameters follow the MATLAB Global Optimization Toolbox defaults \cite{globalOptimizationToolbox}. While GAs are stochastic optimizers with empirical convergence behavior, running multiple restarts improves robustness and reduces the risk of poor local minima. To this aim, we perform 10 independent GA runs with different random seeds and report the best (and median) costs in \Cref{sec:results}. Unlike many identification methods used to derive state-space models from input/output signals \cite{sysIdenToolbox}, GA--DMDc does not require the user to prescribe the model order. It searches over subsets directly and thus \emph{implicitly} selects the number of states. The user does not need to run sensitivity sweeps over model order and focuses evaluation on predictive performance. GA--DMDc only requires the user to specify an upper bound on the allowable state count, and the GA returns a compact set that minimizes \(\mathcal{J}\).

\begin{table}
    \begin{center}
    \caption{Key parameters of the GA-based solver. MATLAB Global Optimization Toolbox defaults are used unless noted.}
    \label{tab:GA_parameters}
    \begin{tabular}{| c | c | c |}
    \hline
    \textbf{GA parameter} & \textbf{Description} & \textbf{Value} \\
    \hline
    \multirow{2}{*}{Population Size} 
      & Chosen from sweep for accuracy  & \multirow{2}{*}{480} \\ 
      & vs.\ runtime                    & \\
    \hline
    Max Number of 
      & Default, \(100 \times\) number  & \multirow{2}{*}{54500} \\
    Generations 
      & of variables                    & \\
    \hline
    Elite Count 
      & Default, \(0.05 \times\) population size & 24 \\
    \hline
    Crossover Fraction 
      & Default                         & 0.8 \\
    \hline
    \end{tabular}
    \end{center}
\end{table}

\section{Recursive Feature Elimination based Algorithm for State Variable Selection}
\label{sec:RFE_DMDc}
In this work, we adapt the RFE algorithm to the selection of a small, physically interpretable subset of measured variables that can serve as states in a DMDc model. The method blends RFE with a model-based importance score derived from the DMDc output map and adds a \emph{cross-subsystem} correction to mitigate feature overshadowing in multi-component systems. We use GA--DMDc (\Cref{sec:GA_DMDc}) as a baseline for validation and compute–accuracy comparisons.

\subsection{Pre-filtering stage}
\label{sec:preFilteringStage}
Before RFE, we automatically exclude candidate variables that are unlikely to contribute useful dynamic information:
\begin{enumerate}
  \item \textit{Input-collinear channels:} any process variable whose absolute correlation with an input channel exceeds a certain threshold is discarded.
  \item \textit{Near-constants:} channels with negligible variability over the training transients, e.g., \(\mathrm{Var}(z_i)<\epsilon\), are dropped. 
  \item \textit{Duplicates/aliases (optional):} highly collinear pairs among process variables can be clustered (e.g., hierarchical clustering) and reduced to representatives to improve conditioning.
\end{enumerate}
This automatic pruning reduces the candidate pool and avoids wasting RFE iterations on trivially uninformative channels.

\subsection{RFE with a DMDc-based importance score}
\label{sec:RFEBasedSelection}
RFE proceeds as backward elimination over the candidate pool \(\mathcal{Z}\):
\begin{enumerate}
  \item \textbf{Evaluation model.} Train a DMDc model on the current feature set to obtain \(\hat A^{d},\hat B^{d},\hat C^{d}\) matrices.
  \item \textbf{Per-variable importance.} Use \(\hat C^{d}\) to quantify how each state affects each output, then aggregate across outputs to score variables.
  \item \textbf{Eliminate and iterate.} Remove the lowest-scoring variables (one or a block), retrain, and repeat until the target cardinality is reached.
\end{enumerate}

The base version of RFE removes one feature per iteration until the target size is reached. Because each step requires refitting an evaluation (or correlation/sensitivity) model to re-rank the remaining candidates, we accelerate the process by pruning multiple low-scored features per iteration. This yields a nested subset chain \(F_1 \subset F_2 \subset \cdots \subset F\), improving efficiency while preserving monotonicity. To further curb ill-conditioning from collinear candidates, we apply a hierarchical clustering prefilter \cite{hierClustering}. The development of an automated state-selection strategy is driven by the observation that \(\hat{C}^d\) acts as a transfer from state space to output space: given snapshots \(X\) and \(Y\), we estimate the output map by least squares (\Cref{eq:dmdc_C_fit}). Since output channels differ in units and gain, we perform \emph{rowwise} min–max scaling of \(\hat{C}^d\) to obtain a dimensionless importance matrix \(I \in \mathbb{R}^{p \times n}\) (\Cref{eq:importanceMetric}); variables with the smallest average importance \(\overline{I}(j)\) are eliminated first.
\begin{equation}
\label{eq:importanceMetric}
I(i,j) \;=\; \frac{\hat C^{d}(i,j) - \min_{j'}\,\hat C^{d}(i,j')}{\max_{j'}\,\hat C^{d}(i,j')-\min_{j'}\,\hat C^{d}(i,j')}\,
\end{equation}
where \(i=1,\ldots,p\) indexes outputs and \(j=1,\ldots,n\) indexes candidate states. We then define a scalar score per variable by averaging across outputs (\Cref{eq:expectedImportanceMetric}).
\begin{equation}
\label{eq:expectedImportanceMetric}
\overline{I}(j)= \frac{1}{p}\sum_{i=1}^{p} I(i,j)
\end{equation}

\begin{figure}[htbp]
    \centering        \includegraphics[width=0.55\textwidth,clip,keepaspectratio]{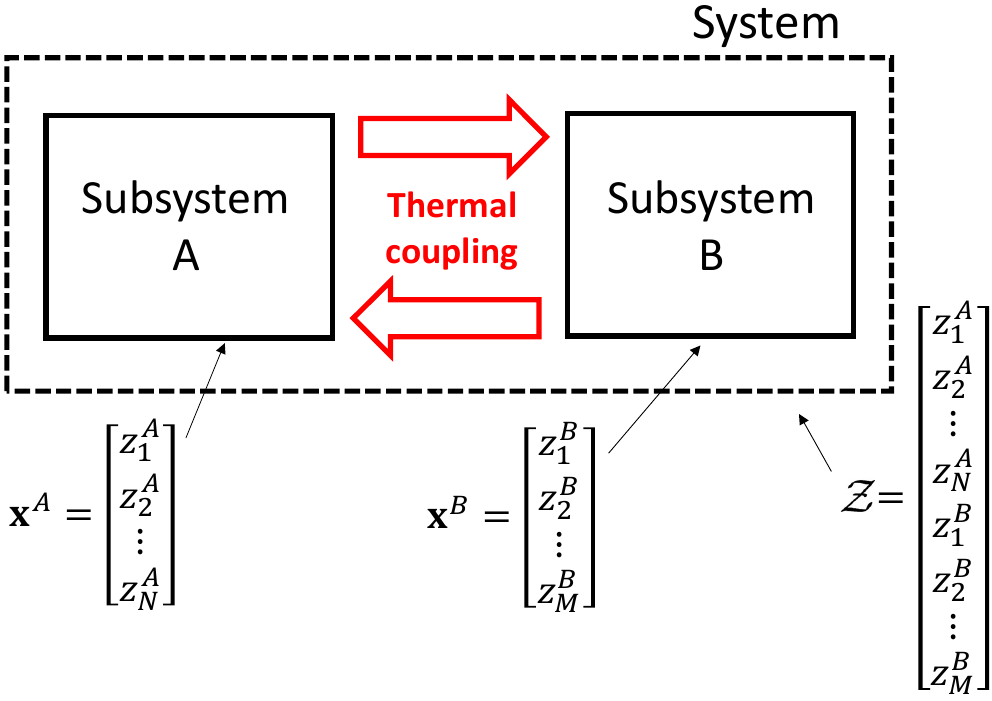}
    \caption{Subsystems interacting on each other with corresponding sets of process variables.}
    \label{fig:Subsystems}
\end{figure}

Let us consider a dynamic system constituted by two subsystems, $A$ and $B$ (\Cref{fig:Subsystems}). Let us assume that $N$ process variables ($\mathbf{z}^A$) are recorded for subsystem $A$, and $M$ process variables ($\mathbf{z}^B$) are recorded for subsystem $B$. The goal is to select the state variables of the coupled system, i.e., the minimum set of $(n+m)$ process variables describing its dynamics ($\mathbf{z}^{opt}\equiv \mathbf{x}^{System}$), where $n<N$ and $m<M$. 
In multi-component systems, naive row-wise normalization of \(\hat C^{d}\) can bias rankings toward variables from a high-gain subsystem: if one subsystem consistently produces larger \(|\hat C^{d}|\) entries, its columns dominate the min–max scaling and inflate \(\overline{I}(j)\) relative to other subsystems. This phenomenon was named feature \textit{overshadowing}. It was observed empirically when applying RFE to the full candidate set. Initially, the RFE-based scheme was applied to the entire set of process variables, without discriminating between the subsystems. The results were not satisfactory, i.e., most of the selected variables belonged to a single, high-impact subsystem. To explain the issue, a not-rigorous demonstration and a numerical example are provided. Let us recast $\hat{C}^d$ matrix in its explicit form, where all the matrix elements representing the impact of the state variables on the output variables are shown. The $\hat{C}^d$ matrix for a system characterized by $p$ output variables, and $n$ state variables is reported in \Cref{eq:CexplicitMatrix}. 
\begin{equation}
\label{eq:CexplicitMatrix}
C(y_1,\cdots,y_p,x_1,\cdots,x_n)=
\begin{bmatrix}
  \frac{\partial y_{1}}{\partial x_1} & \cdots & \frac{\partial y_{1}}{\partial x_n} \\
  \vdots & \ddots & \vdots \\
  \frac{\partial y_{p}}{\partial x_1} & \cdots & \frac{\partial y_{p}}{\partial x_n}
\end{bmatrix}
\end{equation}
By using \Cref{eq:importanceMetric} for calculating the importance matrix, each term of the above matrix is then scaled using the formula expressed by \Cref{eq:CexplicitMatrixScaled}.
\begin{equation}
\label{eq:CexplicitMatrixScaled}
I(i,j) \;=\; \frac{\frac{\partial y_{i}}{\partial x_j} - \min_{j'}\,\hat C^{d}(i,j')}{\max_{j'}\,\hat C^{d}(i,j')-\min_{j'}\,\hat C^{d}(i,j')}\,
\end{equation}

Now let us consider the $\hat{C}^d$ matrix for a system constituted by two subsystems (see \Cref{fig:Subsystems}), each with a single output variable ($p=2$). By using \Cref{eq:CexplicitMatrixScaled} and imposing the conditions reported in \Cref{eq:conditionsExample}, the results in \Cref{eq:conditionsExampleScaled} are obtained.
\begin{equation}
\label{eq:conditionsExample}
C(1,j) > C(2,k), \text{ } \forall j \in [1,n/2],  \forall k  \in (n/2,n] 
\end{equation}
\begin{equation}
\label{eq:conditionsExampleScaled}
\begin{matrix}
I(1,j) > I(2,k), \text{ } \forall j \in (1,n/2],  \forall k  \in (n/2+1,n] \\  \text{ } \\
\max(I(1,:)) = \max(I(2,:)) = 1,  \min(I(1,:)) = \min(I(2,:)) = 0
\end{matrix}
\end{equation}

Each term in $I(2,:)$ is smaller than those in $I(1,:)$ except for the variables corresponding to the minimum and maximum in each row. Such a trend causes most of Subsystem 1 variables to be prioritized with respect to Subsystem 2 variables. The reason can be intuitively understood through a numerical example. Let us consider the system shown in \Cref{fig:Subsystems}, and let us assume that $N=3$ and $M=3$, the conditions reported in \Cref{eq:conditionsExample} for $\hat{C}^d$ matrix are met and each subsystem has one output variables ($p=2$). In \Cref{tab:CmatrixShadowing}, the $\hat{C}^d$ matrix is reported. In \Cref{tab:importanceRankingShadowing}, the importance scores and their expected values computed by using the formulation in \Cref{eq:importanceMetric} and \Cref{eq:expectedImportanceMetric} are reported.

\begin{table}[htb!]
\centering
\caption{$\hat{C}^d$ matrix for a system composed by two subsystems.}
\label{tab:CmatrixShadowing}
\begin{tabular}{|c|c|c|c|c|c|c|}
\hline
\multicolumn{1}{|l|}{} & $\bf z_{1}^{A}$ & $\bf z_{2}^{A}$ & $\bf z_{3}^{A}$ & $\bf z_{1}^{B}$ & $\bf z_{2}^{B}$ & $\bf z_{3}^{B}$ \\ \hline
$y_{1}^{A}$            & 1.00E+2 & 1.00E+1 & 1.00E0 & 1.00E-4 & 1.00E-4 & 1.00E-4 \\ \hline
$y_{1}^{B}$            & 1.00E-5 & 1.00E-5 & 1.00E-5 & 1.00E-1 & 1.00E-3 & 1.00E-4 \\ \hline
\end{tabular}
\end{table}

\begin{table}[htb!]
\centering
\caption{Feature ranking for a system composed by two subsystems, affected by importance overshadowing.}
\label{tab:importanceRankingShadowing}
\begin{tabular}{|c|c|c|c|c|c|c|}
\hline
               & $\bf z_{1}^{A}$ & $\bf z_{2}^{A}$ & $\bf z_{3}^{A}$ & $\bf z_{1}^{B}$ & $\bf z_{2}^{B}$ & $\bf z_{3}^{B}$ \\ \hline
$I^{A}$        & 1.00E0 & 1.00E-1 & 1.00E-2 & 0.00E0 & 0.00E0 & 0.00E0 \\ \hline
$I^{B}$        & 0.00E0 & 0.00E0 & 0.00E0 & 1.00E0 & 9.90E-3 & 9.00E-4   \\ \hline
$\overline{I}$ & 5.00E-1 & 5.00E-2 & 5.00E-3 & 5.00E-1 & 4.95E-3 & 4.50E-4  \\ \hline
\end{tabular}
\end{table}

\Cref{tab:CmatrixShadowing} shows that variables $z_{i}^{A}$ well describe system $A$ dynamics, but they barely represent system $B$ dynamics ($z_{i}^{B}$ variables show the opposite behavior). \Cref{tab:CmatrixShadowing} also shows that $C(1,j)>C(2,k), \text{ } \forall j\in[1,3],  \forall k \in (3,6]$. The average importance values  $\overline{I}$ reported in \Cref{tab:importanceRankingShadowing} entail that the variables belonging to system $B$, except for the its most important one, are always ranked lower than the ones belonging to system $A$. Accordingly, the state variables belonging to system $A$ are characterized by higher importance scores, and they will be prioritized during the selection process. To ensure an unbiased selection, an alternative importance-ranking methodology was proposed. The main steps are described below.

\begin{itemize}
    \item \textit{Step I — Within-subsystem RFE}\\    
    Apply the RFE procedure independently to each subsystem. At the end of this stage, the two sets of candidate state variables for subsystems $A$ and $B$ ($\mathbf{x}^A$ and $\mathbf{x}^B$) are selected.
    \begin{equation}
    \label{eq:xA_xB}
     \mathbf{x}^A=[x_1^A,\ldots, x_n^A], \mathbf{x}^B=[x_1^B, \ldots, x_m^B]
    \end{equation}
    \item \textit{Step II — Cross influence analysis}\\  
    Quantify cross effects using correlation (or partial correlation) between the recorded variables of \(A\) and the candidate states/outputs of \(B\), and vice versa. This step is crucial even when the subsystems are not interacting since the process variables of a subsystem can be privileged, being characterized by a higher scores through the importance metrics (\Cref{eq:expectedImportanceMetric}). It is worth stressing that the impact on both the candidate state variables and the output variables needs to be accounted for (\Cref{eq:cross_A}, \Cref{eq:cross_B}). As a result, two new sets of candidate state variables are obtained ($\mathbf{x}^{A\rightarrow B}$ and $\mathbf{x}^{B\rightarrow A}$).    
    \begin{align}
    \label{eq:cross_A}
    \mathbf{x}^{A\rightarrow B} &:= \text{Vars in }\mathbf{z}^{A}\text{ most informative for }[\mathbf{x}^{B},\mathbf{y}^{B}],\\
    \label{eq:cross_B}
    \mathbf{x}^{B\rightarrow A} &:= \text{Vars in }\mathbf{z}^{B}\text{ most informative for }[\mathbf{x}^{A},\mathbf{y}^{A}]
    \end{align}
    
    \item \textit{Step III — Cross-subsystem merge and search}\\
    Form the merged pool \(\mathcal{Z}_{\times}=\mathbf{x}^{A}\cup \mathbf{x}^{B}\cup \mathbf{x}^{A\rightarrow B}\cup \mathbf{x}^{B\rightarrow A}\) and run a final elimination/search (RFE or small parallel sweep) to minimize the supervised cost in \Cref{eq:cost_function}, yielding \(\mathbf{z}^{opt}\).\\     
\end{itemize}

The workflow is implemented in the RAVEN framework \cite{rabiti2021raven}  The code is available from GitHub repository \href{https://github.com/idaholab/raven}{https://github.com/idaholab/raven}.

\section{Assessment of the performance of the proposed algorithm}
\label{sec:assessment}
The capabilities of the proposed method will be demonstrated by applying it to two different test cases. In \Cref{sec:toy_problem}, it was tested on a simple example problem. Once collected the set of constitutive equations, the state-space representation model was analytically derived, and the corresponding state variables were identified. At the same time, a dataset containing the response of the process variables to a wide range of input perturbation was built, and supplied to the two data-driven methods. The objective is to evaluate how well the state variables identified by the data-driven approaches align with those predicted by the analytical model, as well as to evaluate the stability of GA--DMDc algorithm that is of stochastic nature. In \Cref{sec:test_case}, the methods were tested on a multi-component system where the capability of the algorithm at accounting for cross-influences between interacting dynamic system is assessed. 

\subsection{Validation on a Simple RLC Benchmark}
\label{sec:toy_problem}
\begin{figure}[htbp]
    \centering        \includegraphics[width=0.45\textwidth,clip,keepaspectratio]{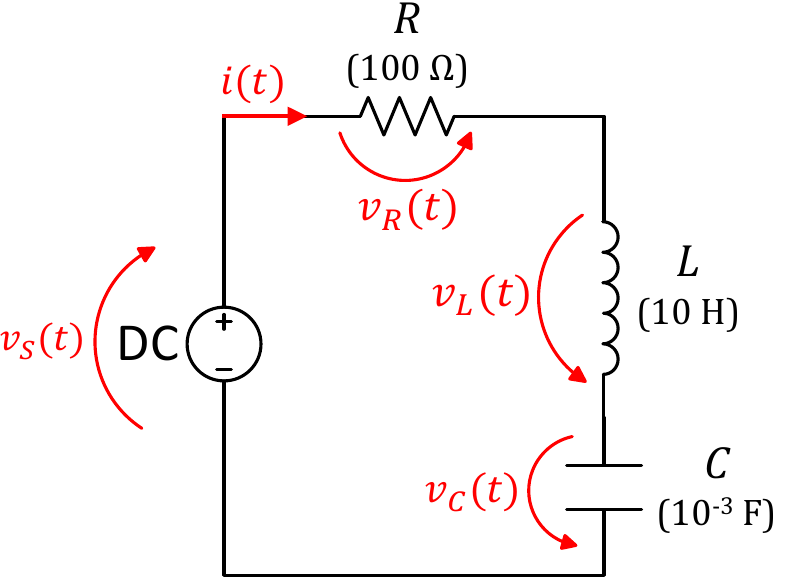}
    \caption{Series RLC circuit used for validation (parameter values shown).}
    \label{fig:RLC}
\end{figure}

An electrical circuit consisting of a resistor, an inductor and a capacitor, connected in series (RLC circuit) was used (\Cref{fig:RLC}) to validate the developed methods. The constitutive equations of the circuit components are reported in \Cref{eq:RLC_eqs}. The current and voltage balance equations typical of a series configuration are reported in \Cref{eq:series_i} and \Cref{eq:series_v}, respectively.
\begin{align}
    \label{eq:RLC_eqs}
    v_R(t) = R \cdot i(t),~v_L(t) = L \frac{di_L(t)}{dt},~i_C(t) = C \frac{dv_C(t)}{dt}
\end{align}
\begin{subequations}
\begin{align}
    i(t) = i_C(t) = i_L(t) \label{eq:series_i} \\
    v_S(t) + v_R(t) + v_L(t) + v_C(t) =0 \label{eq:series_v}
\end{align}
\end{subequations}

In lumped electrical networks, admissible state variables are associated with energy storage: inductors store magnetic energy in the current, and capacitors store electric energy in the voltage. The voltages of the nodes and the currents through components in the circuit are usually the state variables. Thus, for the series RLC in \Cref{eq:RLC_eqs}–\Cref{eq:series_v}, a continuous-time state-space model arises directly by taking the loop current and the capacitor voltage as states (\((v_C,i)\)). We use the source voltage as the input (\Cref{eq:input_vars}), and choose as outputs the capacitor voltage and the resistor voltage (see \Cref{eq:state_vars} and \Cref{eq:output_vars}). Substituting the constitutive laws into this state–output choice yields the continuous-time matrices in \Cref{eq:state_matrices} and \Cref{eq:output_matrices}.
\begin{subequations}
\begin{align}
v=v_S(t) \label{eq:input_vars} \\
\begin{pmatrix}x_1\\x_2\end{pmatrix}=\begin{pmatrix}v_C\\i\end{pmatrix} \label{eq:state_vars} \\    \begin{pmatrix}y_1\\y_2\end{pmatrix}=\begin{pmatrix}v_C\\R\cdot i\end{pmatrix} \label{eq:output_vars}
\end{align}
\end{subequations}
\begin{subequations}
\begin{align}
\begin{pmatrix}\dot{x_1}\\\dot{x_2}\end{pmatrix}=\begin{pmatrix}0&1/C\\-1/L&-R/L\end{pmatrix}\begin{pmatrix}x_1\\x_2\end{pmatrix} + \begin{pmatrix}0\\-1/L\end{pmatrix}v \label{eq:state_matrices} \\
\begin{pmatrix}y_1\\y_2\end{pmatrix}=\begin{pmatrix}1&0\\0&R\end{pmatrix}\begin{pmatrix}x_1\\x_2\end{pmatrix} + \begin{pmatrix}0\\0\end{pmatrix}v \label{eq:output_matrices}
\end{align}
\end{subequations}

\begin{figure}[htbp]
    \centering        \includegraphics[width=0.65\textwidth,clip,keepaspectratio]{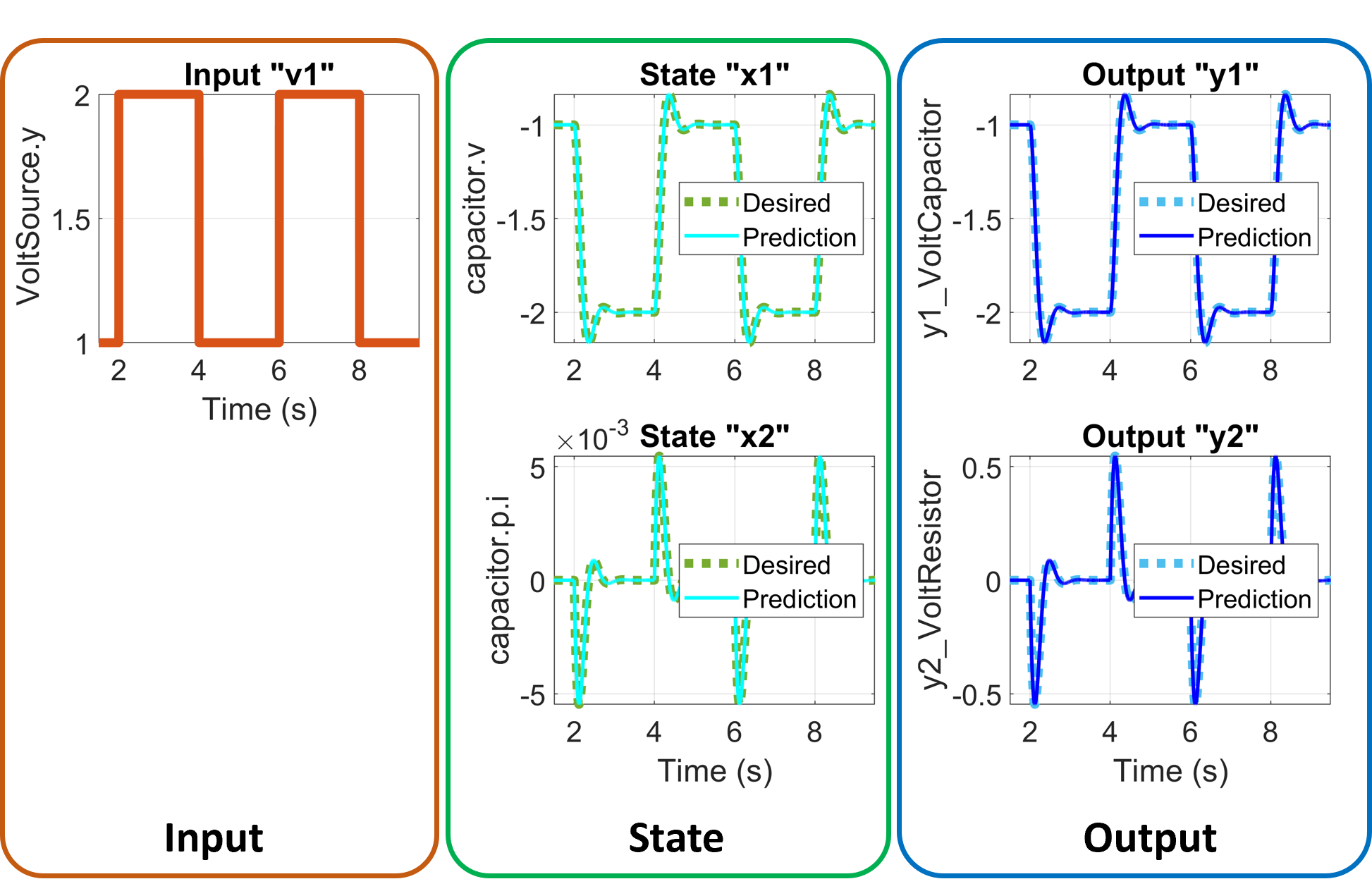}
    \caption{RLC benchmark: predicted state and output trajectories from the data-driven model versus the analytic solution.}
    \label{fig:RLC_Prediction}
\end{figure}

The procedure begins with the creation of a dataset that includes the responses associated with all candidate state variables. We generated an 8\,s dataset (1\,ms sampling) recording 43 process variables, plus the input and two outputs. To span operating conditions, five square-wave excitations with varying offsets and amplitudes were applied to \(v_S(t)\). After the pre-filtering stage (removing constant and highly input-correlated channels), 8 candidates remained and were passed to GA–-DMDc and RFE-–DMDc. With the maximum allowed state count set to 8, both methods independently selected the same two variables: \texttt{capacitor.v} and \texttt{capacitor.p.i}, i.e., the capacitor voltage \(v_C(t)\) and the capacitor current \(i_C(t)\). It can be observed that the performed selection does not correspond to the state variables foreseen by the analytical model \((v_C,i)\). However, the information content is unchanged since the selected variables are 100\% linearly correlated with the analytically predicted ones, i.e., in series, \(i_C(t)\equiv i(t)\) (\Cref{eq:series_i}). To assess robustness, GA–DMDc was run \(10{,}000\) times with different seeds; to avoid enumerating all \(2^8\) masks in a single generation, we temporarily set the population size to 48 (other settings as in \Cref{tab:GA_parameters}). Every run converged to the same two-variable selection, indicating stable convergence and supporting the cost function’s ability to prefer compact, informative state sets. \Cref{fig:RLC_Prediction} compares the predicted state/output trajectories from the learned discrete-time model with the analytic solution.

\subsection{Description of the Multi-component System}
\label{sec:test_case}
\begin{figure}[htbp]
    \centering    
    \includegraphics[width=0.8\textwidth,clip,keepaspectratio]{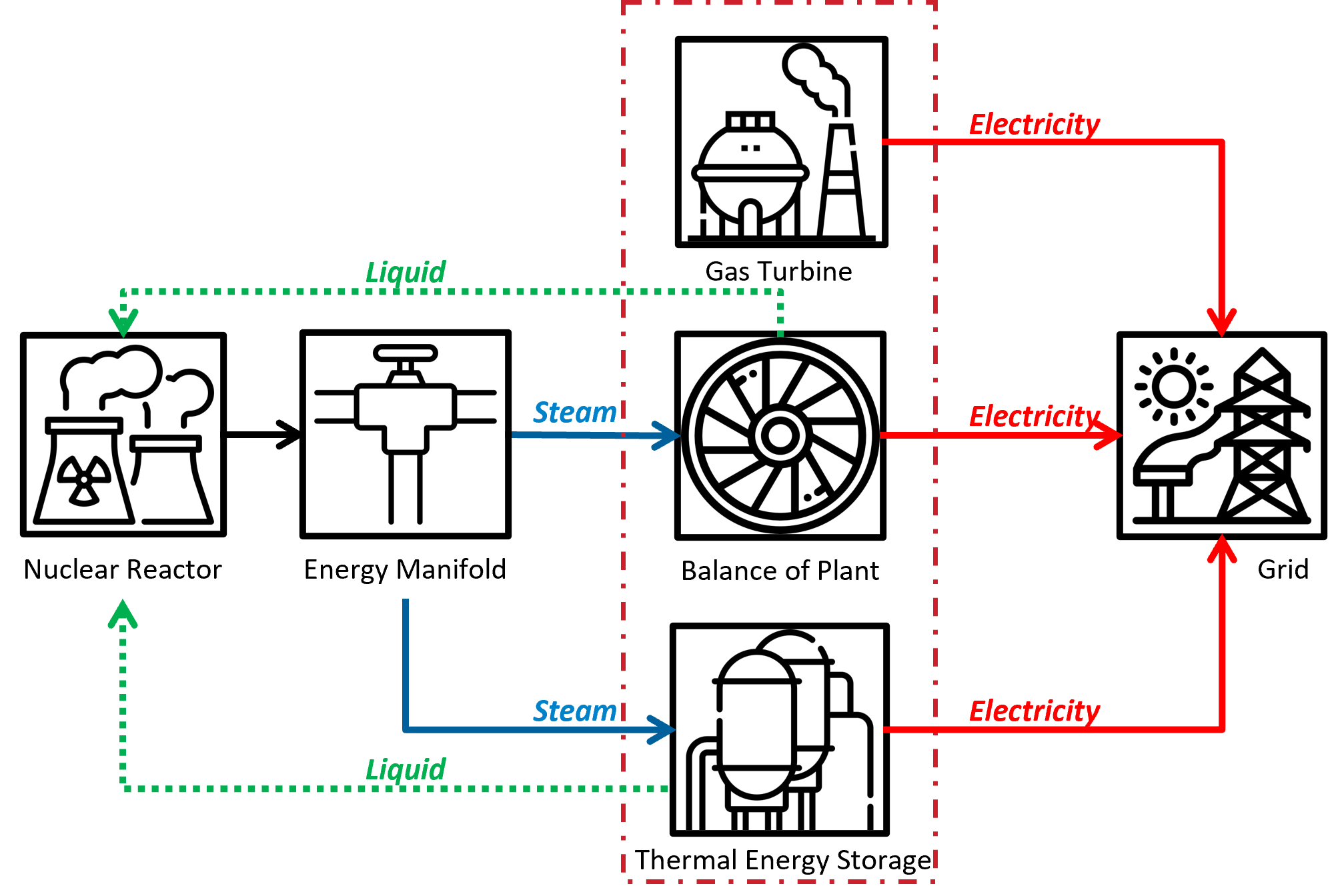}
    \caption{Graphical representation of the IES model\cite{Flaticon}.}
    \label{fig:TriUnits_Dymola}
\end{figure}

To evaluate RFE--DMDc on a realistic Multi-Input Multi-Output (MIMO) setting, we built an Integrated Energy System (IES) in Dymola \cite{systemes2018dymola} using the HYBRID library \cite{HYBRIDrepo} (\Cref{fig:TriUnits_Dymola}). The system comprises three coupled units: a Secondary Energy Source (SES), a Balance of Plant (BOP), and a Thermal Energy Storage (TES). Thermal coupling between BOP and TES is mediated by an Energy Manifold (EM). Because inter-component interactions can bias variable selection, we treat all recorded (nontrivial) process variables as candidate states after statistical pre-processing.

\begin{figure}[htbp]
    \centering    
    \includegraphics[width=0.8\textwidth,clip,keepaspectratio]{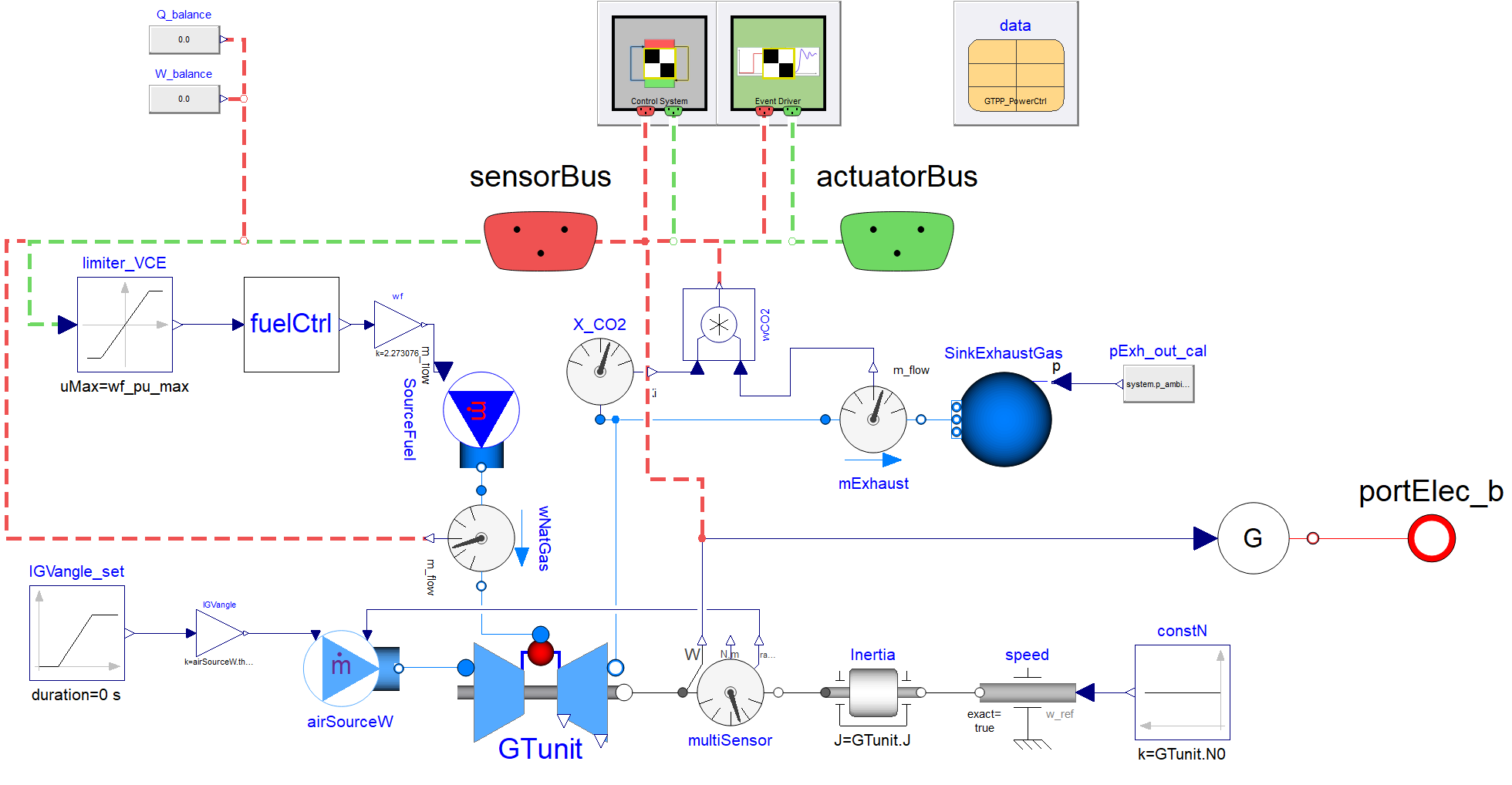}
    \caption{SES Dymola model with sensor and actuator buses.}
    \label{fig:SES_GUI}
\end{figure}

The SES is a 52-MWe natural gas turbine. In \Cref{fig:TriUnits_Dymola}, the SES is the lower icon within the dashed box; its GUI is shown in \Cref{fig:SES_GUI}. Two buses are implemented: \emph{sensorBus} (red dashed lines) aggregates monitored variables, and \emph{actuatorBus} (green dashed lines) carries actuation signals. A PID controller (block \emph{Control}) modulates the gas flow rate in response to the demanded power output variations.

\begin{figure}[htbp]
    \centering    
    \includegraphics[width=0.8\textwidth,clip,keepaspectratio]{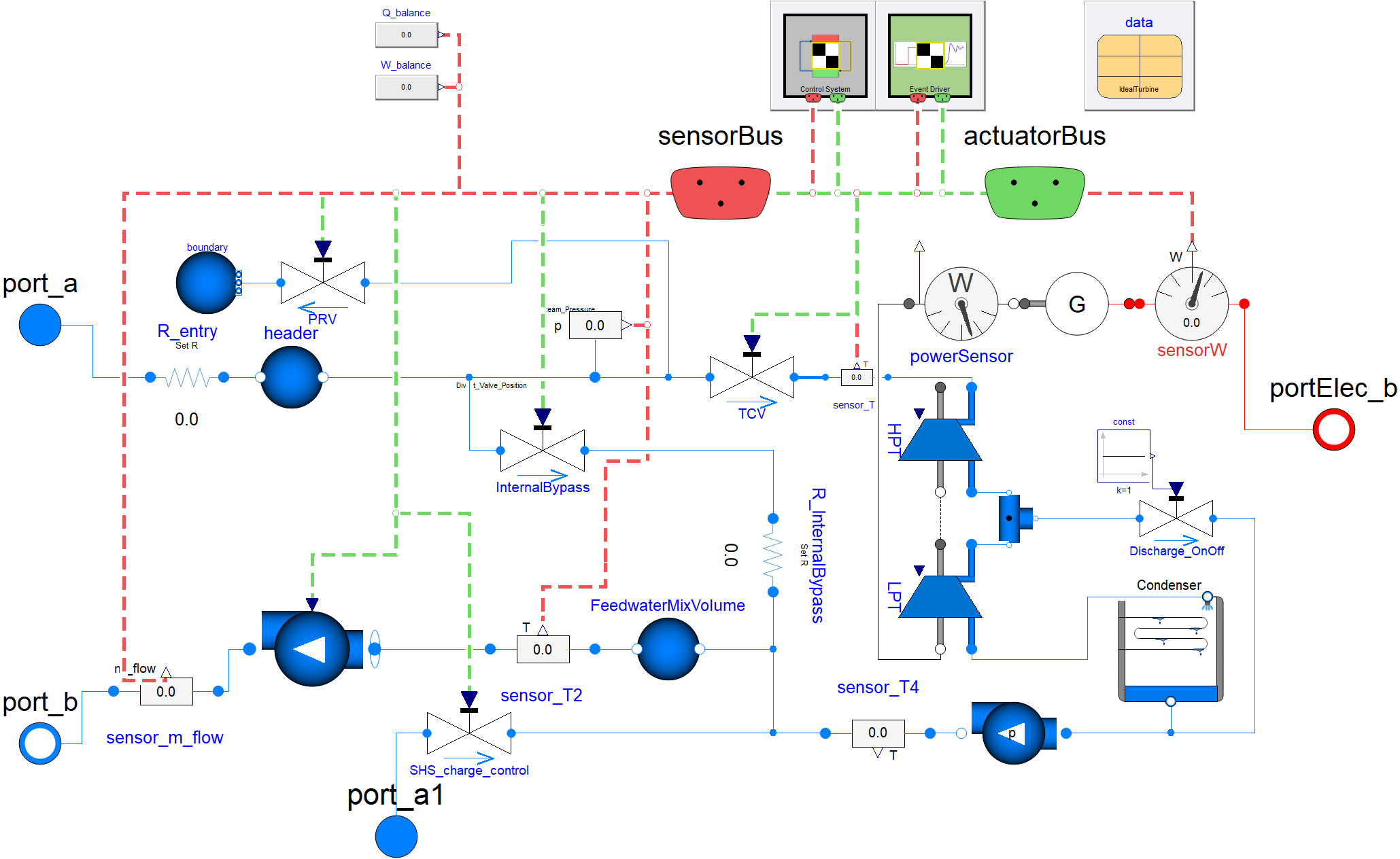}
    \caption{BOP Dymola model (HP/LP turbines, valves, and heat-recovery paths).}
    \label{fig:BOP_GUI}
\end{figure}

The BOP is a 47-MWe Rankine energy conversion cycle. The BOP (upper icon in \Cref{fig:TriUnits_Dymola}) is detailed in \Cref{fig:BOP_GUI}. Similarly to the SES model, measured variables (e.g., inlet steam pressure/temperature, generated electrical power, feedwater temperature and mass flow) feed PID controllers through \emph{sensorBus}; actuators (TCV, InternalBypass, SHS\_charge\_control, feedwater pump) are driven via \emph{actuatorBus}. A portion of the steam flow rate is first fed into the high pressure turbine (HPT), then into the low pressure turbine (LPT) for electric power generation, and finally discharged to the condenser. A fraction is extracted to heat condensate and auxiliary feedwater, which exits at \emph{port\_b}.

\begin{figure}[htbp]
    \centering    
    \includegraphics[width=0.8\textwidth,clip,keepaspectratio]{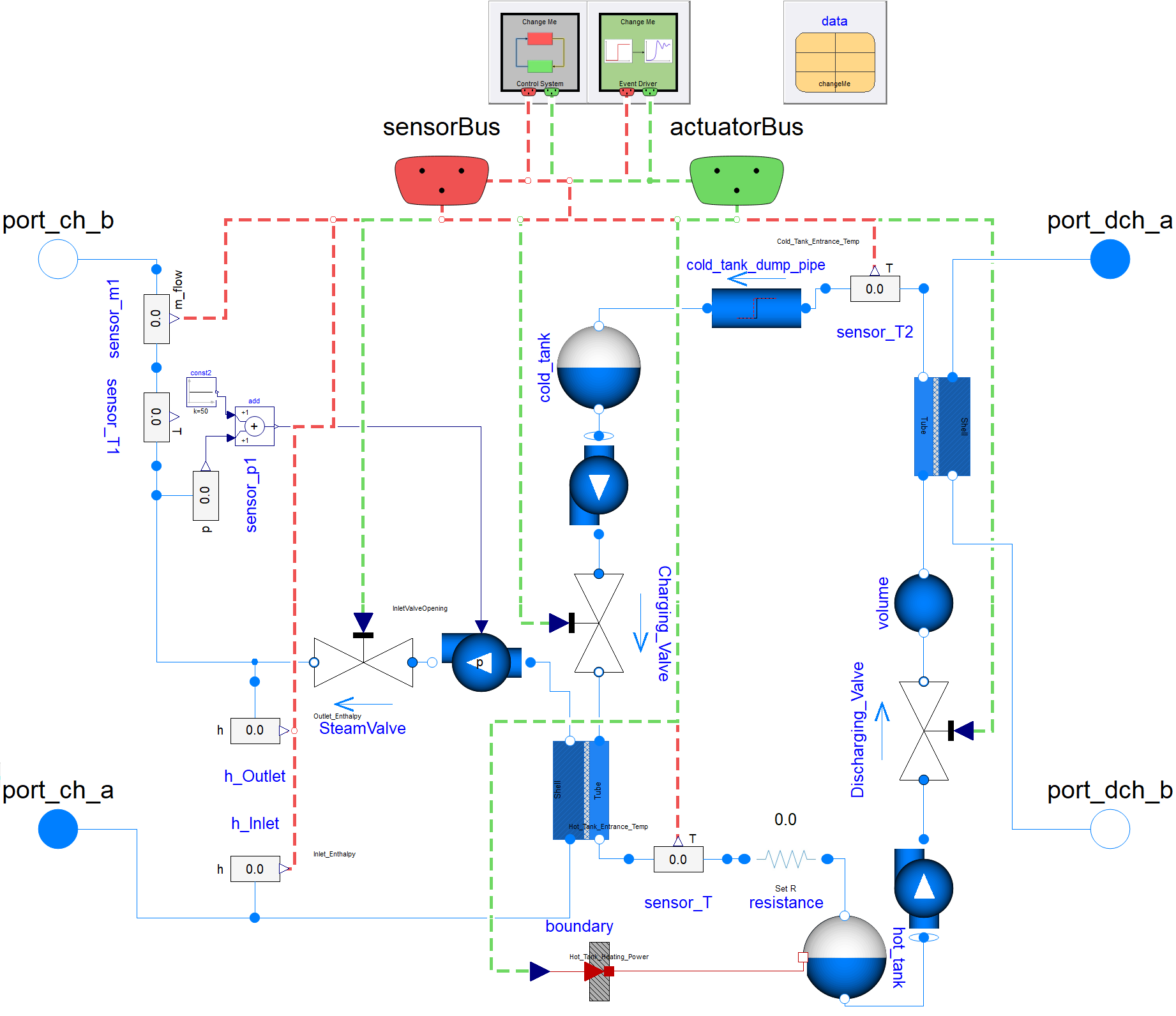}
    \caption{TES Dymola model (two-tank storage, IHX, OTSG, TES turbine).}
    \label{fig:TES_GUI}
\end{figure}

TES consists of hot/cold tanks, two pumps circulating the working fluid (Therminol-66) between the tanks, an intermediate heat exchanger (IHX), and a steam turbine (\Cref{fig:TES_GUI}). In \Cref{fig:TriUnits_Dymola}, the TES is represented by the two, central icons in the dashed-line enclosed area. When the TES is operated in \textit{Charging} mode, the working fluid is heated up by the steam portion that is directed to the TES. After being pumped from the cold tank, the fluid passes through the tube side of the IHX, and then it is stored in the hot tank. The flow rate is controlled by a PID controller to raise the temperature to 240\textdegree C. The steam condensate is collected at the IHX outlet and pumped out of TES via drains. The peak value of the energy storing rate is equal to 4.4 MWe-equivalent. When the TES is operated in \textit{Discharging} mode, the working fluid is pumped from the hot tank to the cold tank through the tube side of a once-through steam generator (OTSG), and water is pumped through the shell side of OTSG to produce a saturated vapor flow rate. The steam is fed to the TES turbine. The peak value of the electrical power output is equal to 14 MWe.

\begin{figure}[htbp]
    \centering    
    \includegraphics[width=0.65\textwidth,clip,keepaspectratio]{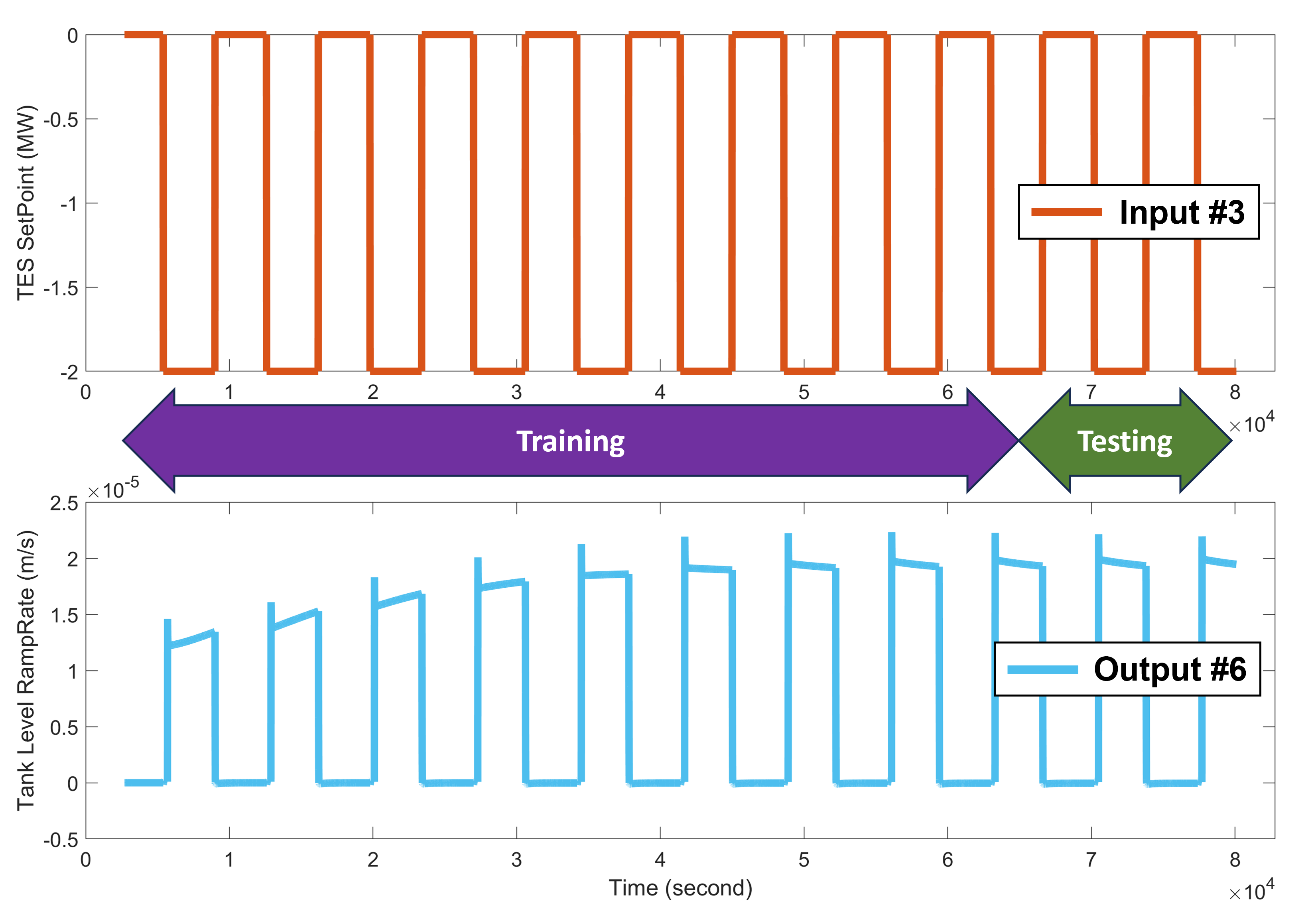}
    \caption{Example input (TES power set-point) and output (hot-tank level ramp rate). Vertical marker denotes the train/test split; early-time drift reflects cold start-up effects.}
    \label{fig:Data_Cold_Start}
\end{figure}

As shown in \Cref{fig:TriUnits_Dymola}, the selected test case receives as inputs the power set-points for the three subsystems, and returns as outputs six process variables. For demonstration purpose, the authors identified two output variables for each subsystem. Regarding the SES, the electrical power output and the firing temperature are selected. These variables need to be monitored to avoid inefficient combustion or excessive thermo-mechanical loads to the turbine blades. Regarding the BOP, the electrical power output and the pressure at the steam turbine inlet are selected, where the latter needs to be monitored to avoid mechanical degradation caused by excessive pressure variations. Regarding the TES, the hot tank filling level and the corresponding rate of variation are selected. The former needs to be monitored to avoid overcharge or depletion of the tanks, and the latter to prevent mechanical damages to the IHX and to the pumps. The list of inputs and outputs is provided in \Cref{tab:IOsummary}.
\begin{table}[htbp]
\centering
\caption{Inputs and outputs of IES unit subsystems.}
\label{tab:IOsummary}
\begin{tabular}{|c|c|c|}
\hline
\textbf{Subsystem} & \textbf{Input} & \textbf{Outputs} \\
\hline
SES & Power set-point & Generated electrical power; Firing temperature \\
\hline
BOP & Power set-point & Generated electrical power; Turbine inlet pressure \\
\hline
TES & Power set-point & Hot-tank level; Hot-tank level ramp rate \\
\hline
\end{tabular}
\end{table}
To select process variables that exhaustively describe the system dynamics, the response of the IES unit was simulated over a wide range of operating conditions. To span operating regimes, each subsystem’s capacity was partitioned into segments (SES: 3; BOP: 4; TES: 6), yielding \(3\times4\times6=72\) operating windows (\Cref{tab:InputTransients}). For each operating window we excite the three inputs with square-wave set-point steps between the window’s lower/upper bounds were applied (period 2\,h, duty cycle \(50\%\)), producing 21.5\,h trajectories per subsystem. During each simulation, the phases of the three square waves were staggered by 900~s to avoid near-collinearity among input channels. Without this staggering, the augmented snapshot matrix
\(\Omega\) in \Cref{eq:dmdc_eq_3} can become (nearly) rank-deficient, which in turn leads to numerical failure or instability in the SVD used to form \(\Omega^{\dagger}\) (\Cref{eq:dmdc_eq_5}). Phase-offset inputs preserve persistent excitation across subsystems, improve the condition number of \(\Omega\), and thereby yield more reliable estimates of \(A^{d}\) and \(B^{d}\).
\begin{table}[htbp]
\centering
\caption{Power-range segments used to define excitation windows.}
\label{tab:InputTransients}
\begin{tabular}{|c|c|c|}
\hline
\textbf{Subsystem} & \textbf{Lower (MWe)} & \textbf{Upper (MWe)} \\
\hline
SES & 13.0 & 26.0 \\ \cline{2-3}
    & 26.0 & 39.0 \\ \cline{2-3}
    & 39.0 & 52.0 \\
\hline
BOP & 7.0  & 17.0 \\ \cline{2-3}
    & 17.0 & 27.0 \\ \cline{2-3}
    & 27.0 & 37.0 \\ \cline{2-3}
    & 37.0 & 47.0 \\
\hline
TES & $-4.4^{*}$ & 0.0 \\ \cline{2-3}
    & $-3.2^{*}$ & 0.0 \\ \cline{2-3}
    & $-2.0^{*}$ & 0.0 \\ \cline{2-3}
    & 0.0 & 2.0 \\ \cline{2-3}
    & 0.0 & 8.0 \\ \cline{2-3}
    & 0.0 & 14.0 \\
\hline
\end{tabular}

\vspace{3pt}
{\centering\footnotesize $^{*}$Negative values denote TES charging power (MWe equivalent).\par}
\end{table}
For each of the 72 windows, we recorded 7{,}096 process variables at 0.1\,Hz. After excluding the 3 inputs and 6 designated outputs, we removed channels that were constant or perfectly collinear (100\% linear correlation) with another channel via a cross-correlation screen, yielding 545 uncorrelated candidates. This reduces computational cost for both GA--DMDc and RFE--DMDc and mitigates numerical issues. Consistently with the selection criteria mentioned in \Cref{sec:GA_DMDc}, we split each trajectory into \(80\%\) training (first 17.2\,h) and \(20\%\) testing (last 4.3\,h). State variables were selected by processing the training dataset only. The state-space matrices derived from training dataset were used to predict the system response by providing the testing dataset. The results were then compared with the baseline to assess the prediction accuracy. \Cref{fig:Data_Cold_Start} shows an example: Input~\#3 (TES power set-point) and Output~\#6 (hot-tank level ramp rate). Although the input’s period and amplitude are fixed, the response evolves due to cold start-up dynamics. The selected state variables are expected to capture the dynamics for both the cold-start period and steady state conditions represented by the responses of process variables in the training dataset.

\section{Results}
\label{sec:results}
All experiments were executed on a Windows~10 workstation with dual Intel$^{\circledR}$ Xeon$^{\circledR}$~8168 CPUs (2.7\,GHz; 48 cores / 96 threads total). In this section, the unified accuracy cost
\(\mathcal{J}\) from \Cref{eq:cost_function} on both training and testing sets, together with wall-clock times are reported.

\subsection{GA--DMDc: population sweep and baseline}
\begin{figure}[htbp]
    \centering    
    \includegraphics[width=0.5\textwidth,clip,keepaspectratio]{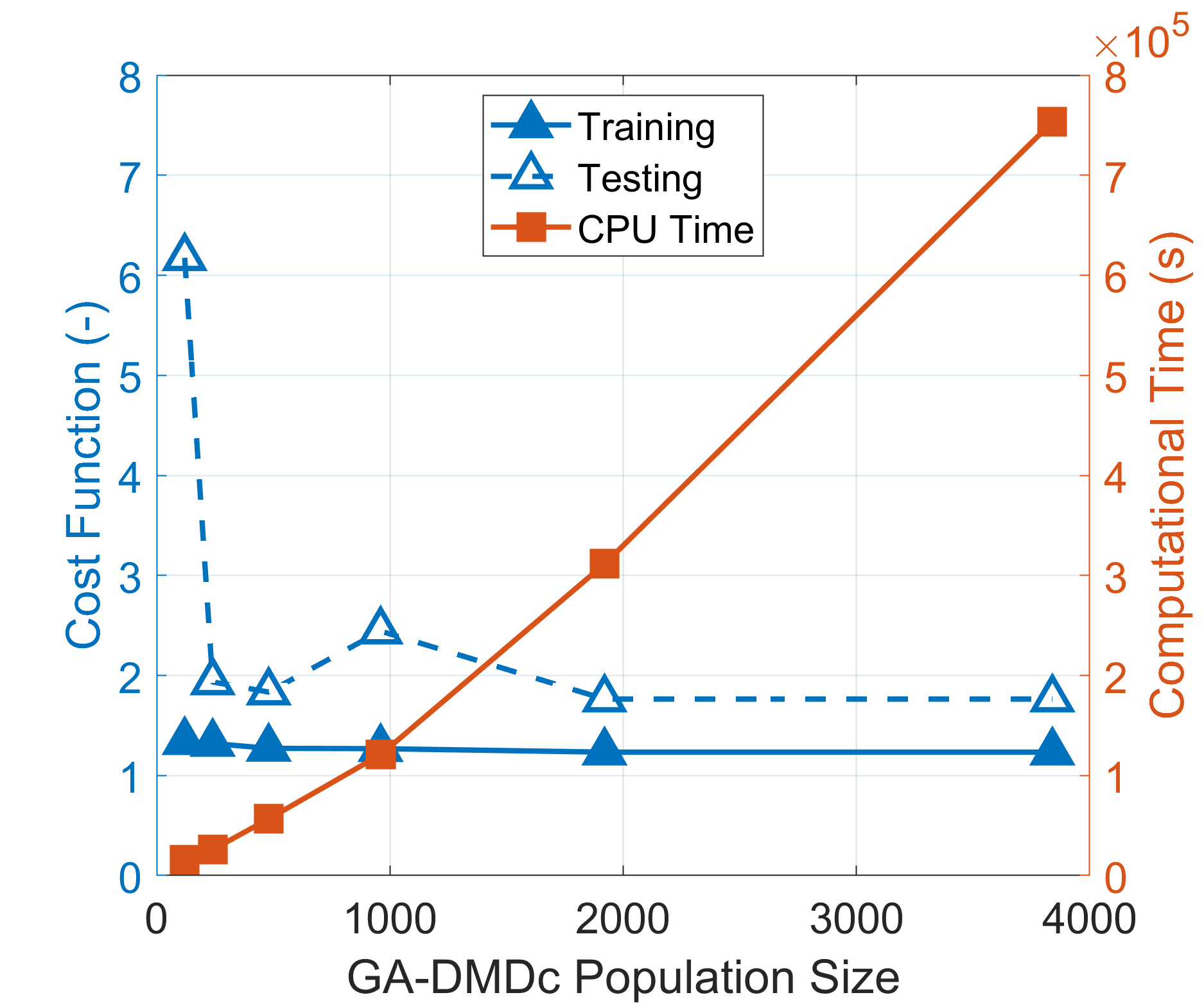}
    \caption{GA--DMDc: cost on training/testing and training runtime versus population size.}
    \label{fig:GA_DMDc_Cost_Time_Population}
\end{figure}

We first study GA--DMDc as a baseline and to set GA hyperparameters. The maximum allowed state count was capped at 18 (a conservative upper bound). Population sizes from 120 to 3{,}840 were tested. \Cref{fig:GA_DMDc_Cost_Time_Population} summarizes the trade-off: larger populations reduce the cost \(\mathcal{J}\) but increase runtime steeply.
Relative to a population of 480, a population of 1{,}920 reduces the cost by only \(3\%\!-\!4\%\) while increasing runtime by \(\sim5.5\times\). To balance accuracy and compute, we fix the GA population at 480 for the remaining experiments (this setting underlies \Cref{sec:Validation}).

\subsection{RFE--DMDc vs.\ GA--DMDc: accuracy and selected size}
\label{sec:Validation}
\begin{figure}[htbp]
    \centering    
    \includegraphics[width=0.5\textwidth,clip,keepaspectratio]{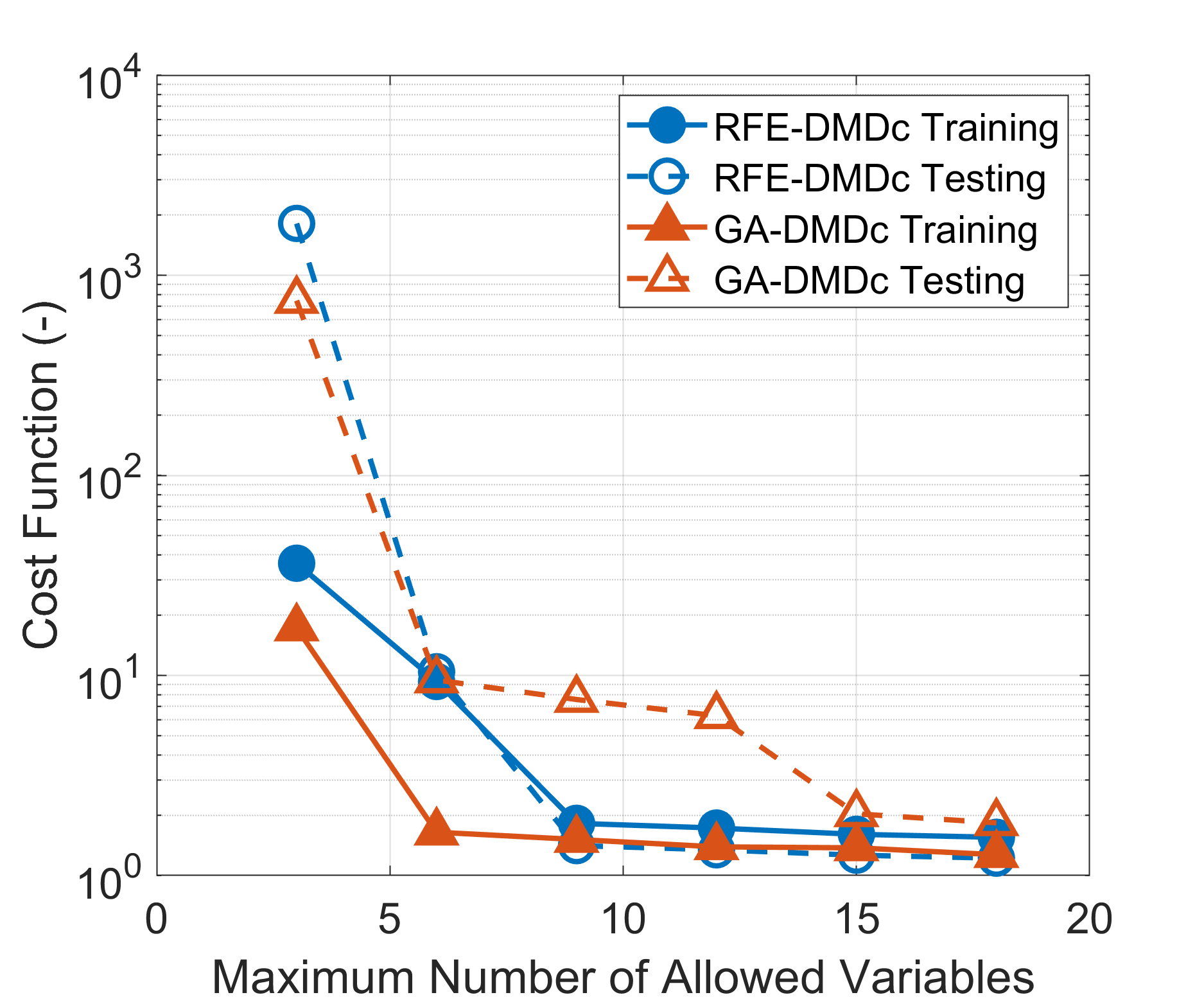}
    \caption{Unified cost \(\mathcal{J}\) for GA--DMDc and RFE--DMDc on training and testing versus the maximum allowed state count.}
    \label{fig:Training_Cost_vs_MaxFeature}
\end{figure}

\begin{figure}[htbp]
    \centering    
    \includegraphics[width=0.5\textwidth,clip,keepaspectratio]{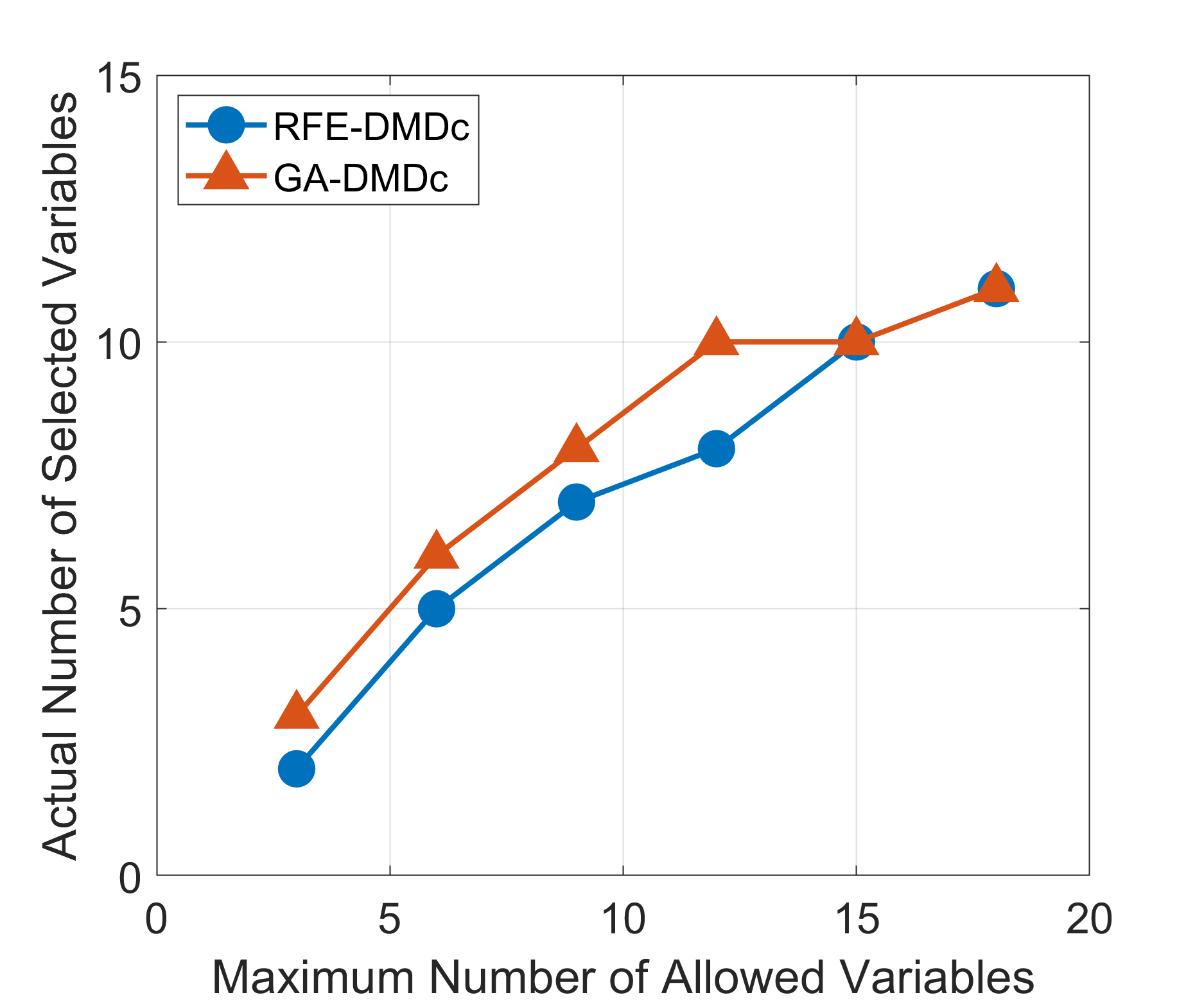}
    \caption{Selected state count versus the maximum allowed count (training). Both methods saturate near \(10\!-\!11\) states.}
    \label{fig:NumFeatures_vs_MaxFeature}
\end{figure}

In this section, the performance of RFE--DMDc and GA--DMDc across a sweep of the \emph{maximum} allowed state count are compared. As mentioned in \Cref{sec:GA_DMDc} and \Cref{sec:RFE_DMDc}, both methods internally choose a subset size up to this cap (i.e., users specify an upper bound, not the model order). Table~\ref{tab:CostFunction_vs_MaxFeatures} reports the achieved costs and the \emph{actual} selected sizes; \Cref{fig:Training_Cost_vs_MaxFeature,fig:NumFeatures_vs_MaxFeature} visualize the trends.

\begin{table}[htbp]
\centering
\caption{Cost and selected size versus maximum allowed variables for RFE--DMDc and GA--DMDc.}
\label{tab:CostFunction_vs_MaxFeatures}
\resizebox{0.82\textwidth}{!}{
\begin{tabular}{|c|cc|c|cc|c|}
\hline
\multirow{2}{*}{\textbf{Max allowed}} & \multicolumn{3}{c|}{\textbf{RFE--DMDc}} & \multicolumn{3}{c|}{\textbf{GA--DMDc}} \\ \cline{2-7}
& \textbf{Train} & \textbf{Test} & \textbf{\# States} & \textbf{Train} & \textbf{Test} & \textbf{\# States} \\
\hline
3  & 36.37 & 1815.68 & 2  & 17.25 & 748.05 & 3  \\ \hline
6  &  9.33 &   10.43 & 5  &  1.65 &   9.49  & 6  \\ \hline
9  &  1.82 &    1.41 & 7  &  1.51 &   7.58  & 8  \\ \hline
12 &  1.73 &    1.34 & 8  &  1.39 &   6.29  & 10 \\ \hline
15 &  1.61 &    1.26 & 10 &  1.38 &   2.03  & 10 \\ \hline
18 &  1.55 &    1.22 & 11 &  1.27 &   1.83  & 11 \\ \hline
\end{tabular}}
\end{table}

From \Cref{fig:Training_Cost_vs_MaxFeature,fig:NumFeatures_vs_MaxFeature}, we note:
\begin{itemize}
  \item Increasing the maximum allowed state count consistently lowers the cost on both training and testing; both methods saturate near \(10\!-\!11\) selected states.
  \item For small caps (3–6), aggressive pruning removes informative channels, which raises the cost \(\mathcal{J}\).
  \item GA--DMDc attains lower \emph{training} cost, whereas RFE--DMDc yields comparable or better \emph{testing} cost—evidence of stronger generalization and reduced overfitting pressure, consistent with RFE’s elimination bias and the cross-subsystem balancing step.
\end{itemize}

\begin{figure}[htbp]
    \centering    
    \includegraphics[width=0.6\textwidth,clip,keepaspectratio]{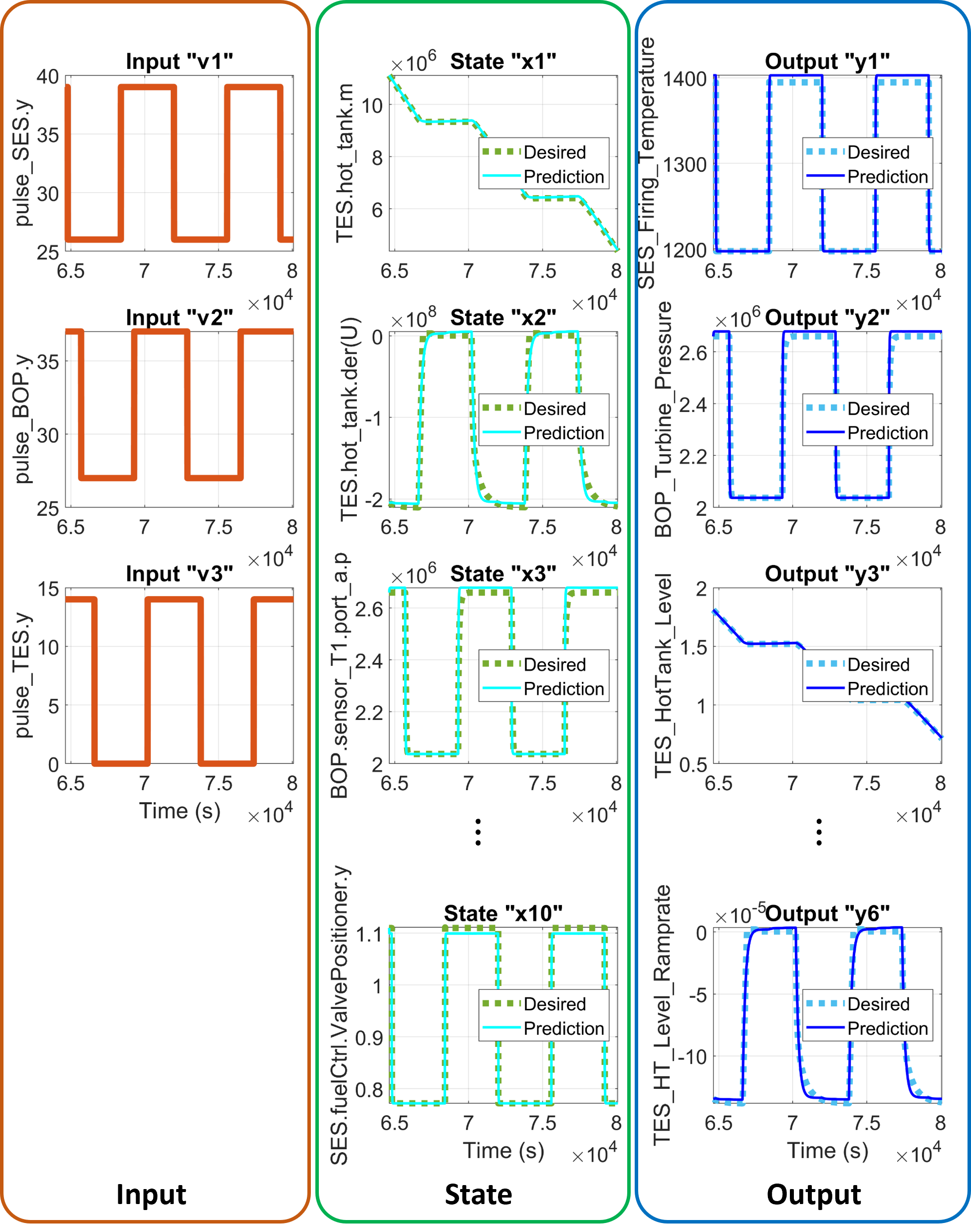}
    \caption{Validation of RFE--DMDc results for a representative run. Predicted values of state and output variables (solid lines) and expected values (dashed lines) are plotted.}
    \label{fig:Predictive_traces}
\end{figure}

\begin{figure}[htbp]
    \centering    
    \includegraphics[width=0.55\textwidth,clip,keepaspectratio]{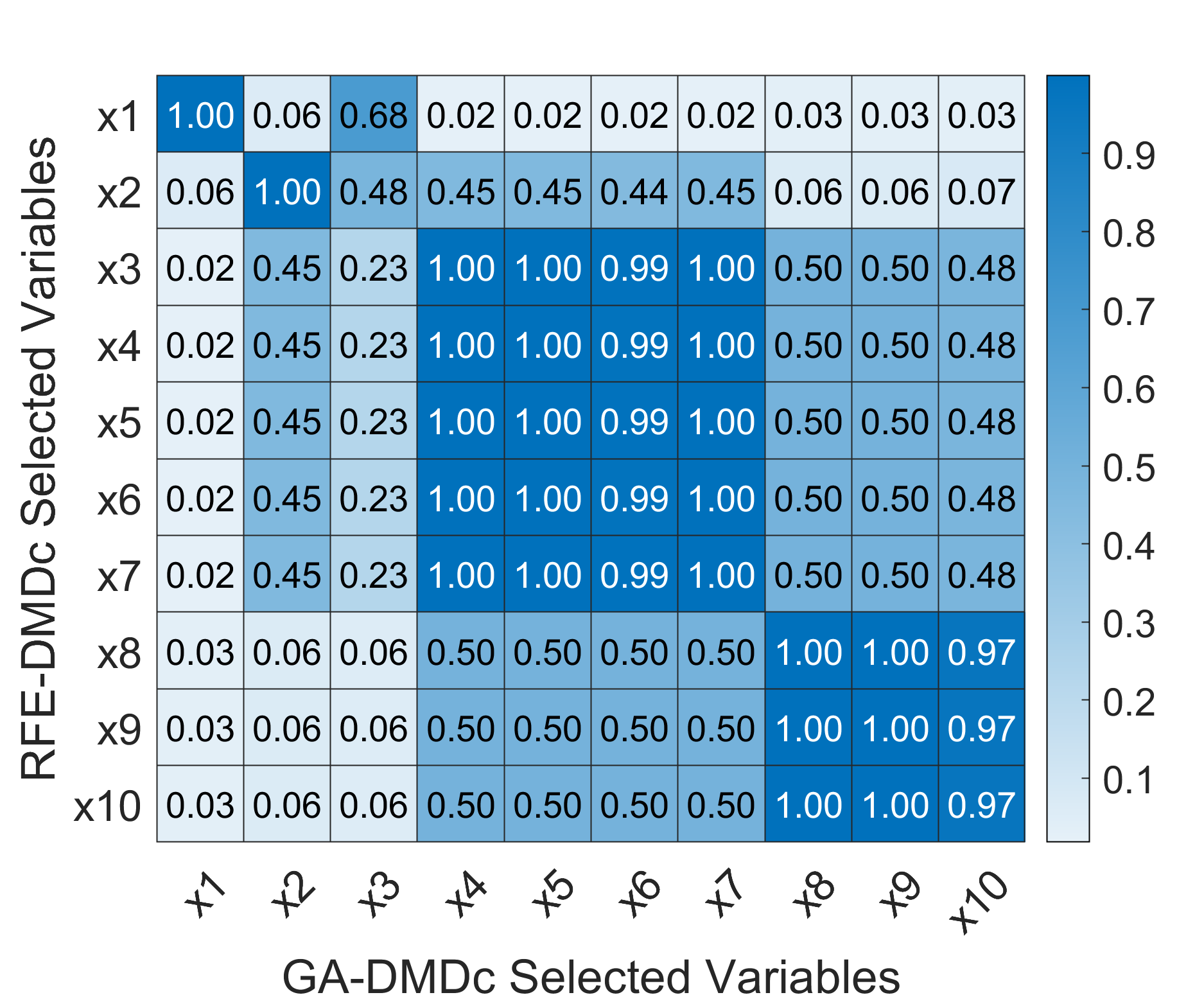}
    \caption{Heat map showing the correlation between the state variables selected by GA--DMDc and RFE--DMDc.}
    \label{fig:correlation_heatmap}
\end{figure}

In \Cref{fig:Predictive_traces}, the outcomes of the selection performed by RFE--DMDc with a max cap of 15 are shown. The method selects 10 states (\Cref{tab:Identity_of_selected_variables}). Predicted (solid) versus reference (dashed) trajectories for states and outputs agree closely. Because the state choice is not unique, the GA--DMDc and RFE--DMDc selections are disjoint (\Cref{tab:Identity_of_selected_variables}), yet deliver similar training costs. A cross-correlation heat map (\Cref{fig:correlation_heatmap}) confirms that the two selected sets encode highly similar information content (numerous coefficients near~1).

\begin{table}
\begin{center}
\caption{Description of the variables selected by RFE--DMDc and GA--DMDc (maximum number of allowed variables set to 15).}
\label{tab:Identity_of_selected_variables}
\begin{tabular}{|c|c|c|}\hline
    \bf \shortstack{State\\Variable} & \bf \shortstack{RFE-DMDc\\Selection} & \bf \shortstack{GA-DMDc\\Selection} \\ \hline
    \multirow{2}{*}{x1} & TES.hot\_tank.m & TES.cold\_tank.V\\
     & (\textit{TES hot tank liquid mass}) & (\textit{TES cold tank liquid volume}) \\ \hline
    
    \multirow{3}{*}{x2} & TES.hot\_tank.der(U) & TES.hot\_tank.der(V)\\
     & (\textit{TES hot tank liquid internal} & (\textit{TES hot tank liquid} \\ 
     & \textit{ energy rate of variation})& \textit{volume rate of variation}) \\ \hline
     
    \multirow{3}{*}{x3} & BOP.sensor\_T1.port\_a.p & TES.discharge\_pump.medium.u\\
     & (\textit{BOP HPT} & (\textit{TES discharge pump} \\ 
     & \textit{inlet pressure}) & \textit{liquid internal energy}) \\ \hline
     
    \multirow{3}{*}{x4} & BOP.CS.sensorBus.Power & BOP.HPT.sat\_in.Tsat\\
     & (\textit{BOP electrical} & (\textit{BOP HPT inlet} \\ 
     & \textit{power output}) & \textit{saturation temperature}) \\ \hline
     
    \multirow{4}{*}{x5} & BOP.CS.sensorBus. & BOP.HPT.dew\_out.d\\
     & Steam\_Temperature & (\textit{BOP HPT outlet} \\
     & (\textit{BOP HPT inlet} & \textit{steam density}) \\ 
     & \textit{steam temperature}) & \\ \hline
     
    \multirow{3}{*}{x6} & BOP.Condenser.port\_a.m\_flow & BOP.HPT.x\_th\_out\\
     & (\textit{BOP condenser inlet} & (\textit{BOP HPT outlet} \\ 
     & \textit{mass flow rate}) & \textit{steam quality}) \\\hline
     
    \multirow{3}{*}{x7} & BOP.sensor\_T1.port\_a.m\_flow & BOP.tee.medium.T\_degC\\
     & (\textit{BOP HPT inlet} & (\textit{BOP LPT inlet} \\ 
     & \textit{mass flow rate}) & \textit{steam temperature}) \\ \hline
     
    \multirow{4}{*}{x8} & SY.CS.sensorBus. & SES.SinkExhaustGas.\\
     & W\_subsystems[3] & medium.d\\
     & (\textit{Power of the third subsystem} & (\textit{SES exhaust gas} \\ 
     & \textit{connected to switch yard}) & \textit{density}) \\ \hline
     
    \multirow{4}{*}{x9} & SES.GTunit.combChamber. & SES.SinkExhaustGas.\\
     & fluegas.T & medium.p\_bar\\
     & (\textit{SES combustion chamber} & (\textit{SES exhaust gas } \\
     & \textit{exhaust gas temperature}) & \textit{pressure}) \\ \hline
     
    \multirow{3}{*}{x10} & SES.fuelCtrl. & SES.fuelCtrl.VCE\\
     & ValvePositioner.y & (\textit{SES fuel valve opening} \\
     & (\textit{SES fuel valve opening}) & \textit{set-point}) \\ \hline
\end{tabular}
\end{center}
\end{table}

\subsection{Computational efficiency of the developed algorithms}
\begin{figure}[htbp]
    \centering    
    \includegraphics[width=0.5\textwidth,clip,keepaspectratio]{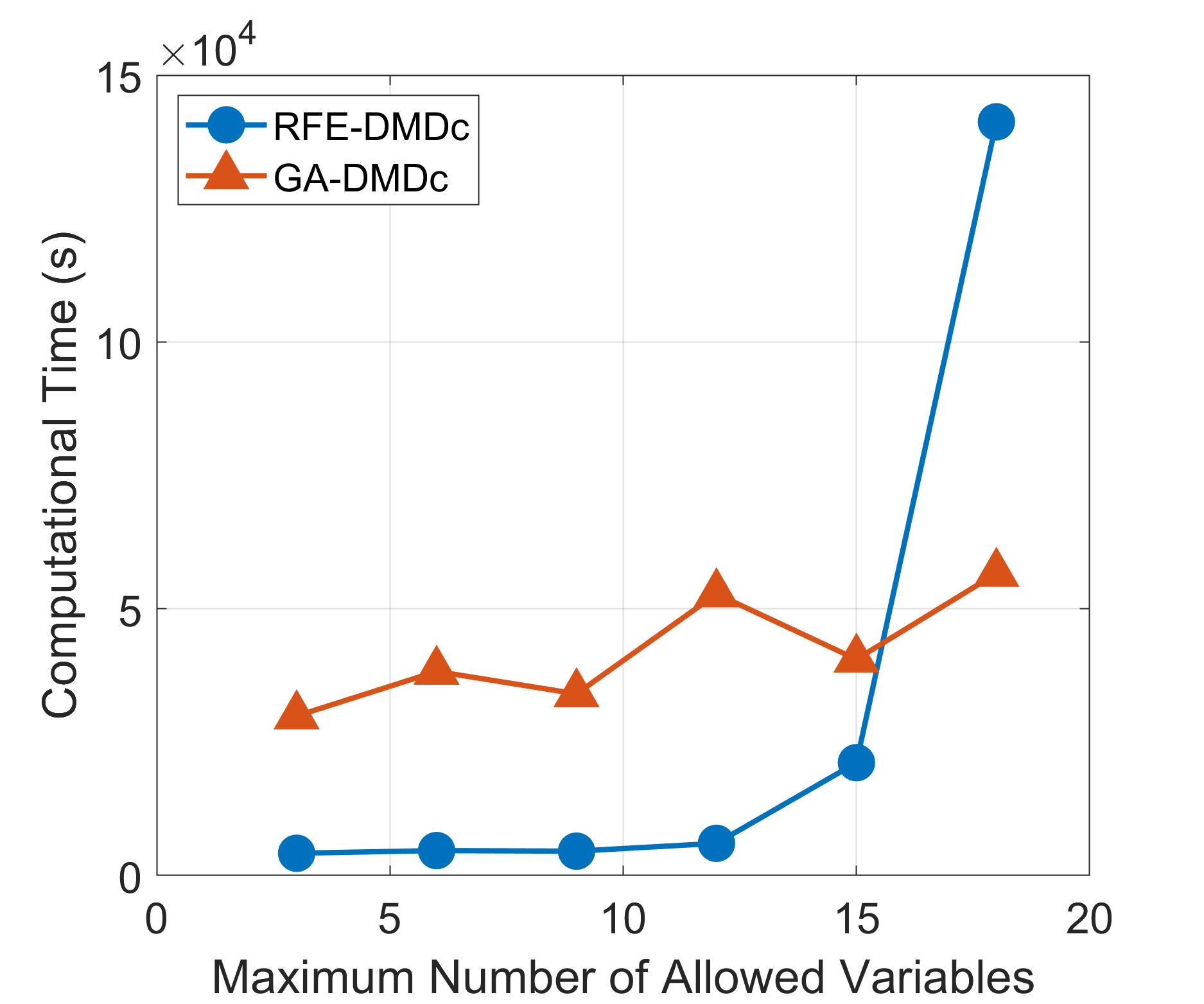}
    \caption{Runtime versus maximum allowed variables. RFE--DMDc shows combinatorial growth from the parallel search stage; GA--DMDc increases with search complexity and stochastic variability.}
    \label{fig:CPU_Time_vs_MaxFeature}
\end{figure}

\Cref{fig:CPU_Time_vs_MaxFeature} reports runtimes versus the maximum allowed state count. For RFE--DMDc, we decompose runtime into four stages: \emph{File I/O}, \emph{Pre-filtering}, \emph{RFE selection}, and \emph{Parallel search}. The last stage dominates as the candidate pool grows, scaling with the number of examined subsets. \emph{File I/O} and \emph{Pre-filtering} are purely sequential and essentially constant for our setup ($\approx$10\,s and $\approx$180\,s, respectively), i.e., independent of the cap on states. \emph{RFE selection} is executed per subsystem and thus parallelizes only across the three subsystems in this IES study; its time is largely insensitive to the cap, up to random fluctuations. During \emph{Parallel search} we evaluate all nonempty cross-subsystem subsets of the $N_S$ shortlisted candidates. Consequently, the parallel time is proportional to the number of the possible combinations $N_{\mathrm{Comb}}$ (\Cref{eq:Ncombine}).
\begin{equation}
    \label{eq:Ncombine}
    N_{\mathrm{Comb}} \;=\; \sum_{j=1}^{N_S} C_{N_S}^j \;=\;  \sum_{j=1}^{N_S}\binom{N_S}{j} \;=\; 2^{N_S}-1    
\end{equation}
where $C_{N_S}^j$ is the number of combinations that can be obtained by selecting $j$ items from a set of $N_S$ distinct members.\\
Ideal efficiency is approached when \(N_{\mathrm{Comb}}\) is much larger than the number of available cores \(P\). \Cref{tab:CPUTime_vs_MaxFeatures} reports the RFE--DMDc breakdown (\emph{RFE selection} and \emph{Parallel search}) together with the number of examined subsets. Fixed overheads $T_0$ are small and predictable; the dominant term is the exhaustive cross-system sweep. Under idealized parallelism the wall time scales as indicated in \Cref{eq:RFE_time_ideal}.
\begin{equation}
\label{eq:RFE_time_ideal}
T_{\mathrm{RFE}} \;\approx\; T_0 \;+\; \frac{2^{N_S}-1}{P}\, t_{\mathrm{fit}} .
\end{equation}
with \(t_{\mathrm{fit}}\) the average DMDc fit/evaluation time per subset.\\
By contrast, GA--DMDc runtime increases with the maximum allowed order (\Cref{fig:CPU_Time_vs_MaxFeature}) because the search space and the number of mutation/crossover pathways grow with the cap.  With the settings in \Cref{tab:GA_parameters}, GA--DMDc employs early stopping upon reaching a local minimum, and thus does not hit any hard cap on the number of iterations. A useful model for the wall time is reported in \Cref{eq:GA_time_ideal}.
\begin{equation}
\label{eq:GA_time_ideal}
T_{\mathrm{GA}} \;\approx\; T_1 \;+\; \frac{R \cdot \mathrm{Pop} \cdot G_{\mathrm{eff}}}{P}\,t_{\mathrm{fit}}
\end{equation}
where \(\mathrm{Pop}\) is the population size, \(G_{\mathrm{eff}}\) the number of generations to convergence, and \(R\) the number of restarts.\\
On the 48-core node, we observe a practical \emph{crossover}: for \(N_S\le 15\), the exhaustive cross-subsystem sweep is faster and deterministic; for \(N_S\ge 16\), the \(2^{N_S}\) growth dominates and GA--DMDc becomes cheaper in wall-clock time (\Cref{tab:CPUTime_vs_MaxFeatures}, \Cref{fig:CPU_Time_vs_MaxFeature}). This threshold shifts with \(t_{\mathrm{fit}}\) (matrix sizes, SVD truncation \(q\)), the available cores \(P\), and I/O overheads.
\begin{table}[htbp]
\begin{center}
\caption{Runtimes versus maximum allowed variables (RFE--DMDc breakdown and GA--DMDc totals).}
\label{tab:CPUTime_vs_MaxFeatures}
\begin{tabular}{|c|c|c|c|c|}
\hline
\textbf{Max.\ Allowed} & \multicolumn{3}{c|}{\textbf{RFE--DMDc}} & \textbf{GA--DMDc} \\ \cline{2-4}
\textbf{Variables} 
& \textbf{RFE Selection} & \textbf{Parallel Search} & \textbf{\# Comb.} & \textbf{Total Time} \\
& \textbf{Time (s)} & \textbf{Time (s)} & \textbf{examined} & \textbf{(s)} \\
\hline
 3  & 3{,}939.96  &     51.42  &        7   & 29{,}902.06 \\ \hline
 6  & 4{,}355.10  &    122.81  &       63   & 38{,}255.21 \\ \hline
 9  & 4{,}028.67  &    357.87  &      511   & 34{,}049.66 \\ \hline
12  & 3{,}791.91  &  2{,}043.15 &    4{,}095 & 52{,}768.92 \\ \hline
15  & 3{,}958.44  & 17{,}003.73 &   32{,}767 & 40{,}529.55 \\ \hline
18  & 4{,}272.59  & 136{,}784.82 & 261{,}971 & 56{,}612.30 \\ 
\hline
\end{tabular}

\vspace{4pt}
\footnotesize\emph{Notes.} “RFE Selection” includes pre-filtering and per-subsystem RFE; “Parallel Search” explores all nonempty cross-subsystem subsets. GA--DMDc column reports end-to-end wall time.
\end{center}
\end{table}

\section{Conclusion}
\label{sec:conclusion}
In this paper, we applied RFE to the state-selection problem for dynamical systems. Given multivariate time series of monitored process variables, the proposed workflow identifies a compact subset that captures the system dynamics and serves as the state for linear surrogate modeling and downstream control/diagnostics. Unlike approaches that learn latent coordinates, the selected variables retain clear physical meaning, aiding interpretation, instrumentation, and validation. The workflow was validated on an IES unit comprising thermally coupled generation and storage subsystems. Using a Dymola model to generate 72 operating windows and time series for 7{,}087 candidates, RFE--DMDc selected small state sets that matched the predictive accuracy of a GA baseline with substantially lower compute. For example, with a cap of 15 candidates, RFE--DMDc identified 10 representative variables in under six hours on a 48-core workstation, and exhibited strong train/test generalization with reduced overfitting. Because the final states are measured variables, the approach naturally supports sensor planning and Digital Twin deployment by prioritizing channels that most improve dynamical prediction with minimal sensing overhead.\\
A key challenge in multi-component settings is feature overshadowing, whereby high-gain subsystems dominate the ranking. Our cross-subsystem step mitigates this effect but induces a combinatorial search. As implemented, the parallel sweep in RFE--DMDc enumerates all nonempty subsets of the \(N_S\) shortlisted variables, so runtime scales as \(O(2^{N_S})\). Empirically, we observe a runtime crossover: beyond \(\sim\!20\) candidates, exhaustive search becomes impractical even with parallelism. Future work will replace exhaustive enumeration with more scalable strategies (e.g., branch-and-bound with admissible bounds, stochastic or greedy subset search with warm starts, and early-pruning heuristics) while preserving the accuracy benefits observed here. In parallel, we will develop global importance metrics that penalize overshadowing without requiring subsystem-wise RFE, thereby removing the need for a priori subsystem partitioning and moving the pipeline toward a fully automated selection procedure.

\section*{Acknowledgments}
This work was supported by the U.S. Department of Energy, Office of Nuclear Energy. The authors specially thank the Integrated Energy System program for collaborative efforts.

%Bibliography
\bibliographystyle{unsrt}  
\bibliography{references}

\end{document}